\documentclass[prd, aps, showpacs]{revtex4}
\usepackage{latexsym, graphicx}

\def\be{\begin{equation}}
\def\te{\end{equation}}
\def\bea{\begin{eqnarray}}

\def\tea{\end{eqnarray}}
\begin{document}

\title{Bose - Einstein Condensate Superfluid-Mott Insulator Transition in an
Optical Lattice}
\author{Esteban Calzetta$^{1}$, B. L. Hu$^{2}$ and Ana Maria Rey$^{2,3,4}$ \\
$^{1}$\textit{\small Departamento de Fisica, Facultad de Ciencias Exactas y
Naturales,}\\
\textit{\small Universidad de Buenos Aires- Ciudad Universitaria, 1428
Buenos Aires, Argentina}\\
$^{2}$\textit{\small Department of Physics, University of Maryland, College
Park, MD 20742}\\
$^3$\textit{\small National Institute of Standards and Technology,
Gaithersburg, MD 20899}\\
$^4$\textit{\small Institute for Theoretical Atomic, Molecular and Optical Physics,}\\
\textit{\small Harvard-Smithsonian Center of Astrophysics,
Cambridge, MA, 02138}}

\date{July 11, 2005}

\begin{abstract}
We present in this paper an analytical model for a cold bosonic gas on an
optical lattice (with densities of the order of $1$ particle per site)
targeting the critical regime of the Bose - Einstein Condensate superfluid -
Mott insulator transition. We focus on the computation of the one - body
density matrix and its Fourier transform, the momentum distribution which is
directly obtainable from `time of flight'' measurements. The expected number
of particles with zero momentum may be identified with the condensate
population, if it is close to the total number of particles. Our main result
is an analytic expression for this observable, interpolating between the
known results valid for the two regimes separately: the standard Bogoliubov
approximation valid in the superfluid regime and the strong coupling
perturbation theory valid in the Mott regime.
\end{abstract}

\maketitle

\section{Introduction}

Since their experimental realization in 1995 \cite{Nobel}, Bose - Einstein
condensates (BEC) have become one of the most exciting fields in physics.
Because the high degree of control and the good understanding of the
microscopic physics involved, they provide an excellent opportunity to
investigate various issues in atomic and molecular physics, quantum optics,
solid state physics and even high energy physics and cosmology \cite
{NatureInsight}.

The interest in these systems is also boosted by its possible use in the
implementation of quantum information processing (QIP)\cite{JZ04}. Cold
neutral atoms in optical lattices are a naturally scalable system, and
because of the weak coupling to the environment long decoherence times are
expected. There are detailed proposals on how to build quantum gates \cite
{MGWRHB03,VSB05} and qubit buses \cite{BSW03} to exchange information
between different locations. All these properties make these systems a
promising candidate for QIP.

In most proposals, the physical qubit is a single atom which may be in one
of two preferred hyperfine states. This implies a strict control of the
number of atoms per site, which in principle may be achieved by driving the
system deep into the Mott insulator (MI) regime \cite{init}. However, the
gas is usually first condensed in a trap, and then the lattice is imprinted
on it. This implies driving the system through the superfluid (SF) -
insulator transition. As with other phase transitions, we expect the
particle distribution will be determined by events at or just below the
critical point; once the hopping parameter is low enough, this distribution
will be simply frozen in \cite{DSZB04}.

To amplify this important point, we observe that it is expected both Landau
and Beliaev damping will be strongly suppressed in the Mott regime \cite
{TG05}; this means that the equilibration times will grow sharply as we
cross from the superfluid to the insulator phases. The pattern of
correlations among different sites and particle number fluctuations will get
frozen once the relaxation time is long compared with the characteristic
time in which the parameters of the model are being changed. Unless this
change is made very slowly, this will happen soon after entering the Mott
regime. In this ``diabatic'' transition, the likelihood of a vacancy or of a
multiply occupied site will correspond to those of a lattice near the
critical point, rather than to the parameters of the operating regime.

The goal of this paper is to formulate an analytical model for a cold
bosonic gas on an optical lattice (with densities of the order of $1$
particle per site) targeting the critical regime of the BEC superfluid -
Mott insulator transition \cite{FWGF89,GMEHB02}. We focus on the computation
of the one - body density matrix \cite{PO56} and its Fourier transform, the
momentum distribution which is directly obtainable from `time of flight''
measurements \cite{GMEHB02,GWFMGB05,GWFMGB05b} (see \cite{RB03}). The
expected number of particles with zero momentum may be identified with the
condensate population, if it is close to the total number of particles. Our
main result is an analytic expression for this observable, interpolating
between the known results valid for the two regimes separately: the standard
Bogoliubov approximation valid in the superfluid regime \cite{am1} and the
strong coupling perturbation theory valid in the Mott regime \cite
{FM94,FM96,EM99,am2}. Comparison of our analytic results with exact
numerical solutions for $N$ particles in a one-dimensional lattice of $N=9$
sites shows that unlike the standard Bogoliubov and strong coupling
perturbation our analytic solution sustains an uniform accuracy throughout.

\subsection{The model}

We consider a system of $N$ particles distributed over $N_{s}$ lattice
sites, with an integer mean occupation number $n=N/N_{s}$. In terms of the
creation and destruction operators $a_{j}^{\dagger }\left( t\right) $ and $%
a_{i}\left( t\right) ,$ the dynamics is described by the Bose - Hubbard
Hamiltonian (BHM) \cite{BC04}

\begin{equation}
H=\sum_{i}\;\left\{ -\sum_{j}J_{ij}a_{i}^{\dagger }a_{j}+\frac{U}{2}%
a_{i}^{\dagger 2}a_{i}^{2}\right\}  \label{nbodyh}
\end{equation}
where the first term describes hopping between sites, and the second term
the in-site repulsion between particles. The matrix $J_{ij}$ is equal to $J$
if the sites $i$ and $j$ are nearest neighbors, and zero otherwise. When the
repulsion term dominates, the ground state of the system has definite
occupation numbers for each site, and weak correlations among different
sites. The system is in the so-called Mott insulator (MI) phase. When the
hopping term dominates, atoms condense into a single quantum state extended
over the whole lattice; the system is in the superfluid (SF) phase.

In this paper we shall focus on the calculation of the one - body density
matrix

\begin{equation}
\sigma _{1lk}=\left\langle a_{l}^{\dagger }\left( t\right) a_{k}\left(
t\right) \right\rangle ,  \label{onebodydenmat}
\end{equation}
and its Fourier transform

\begin{equation}
N_{q}=\frac{1}{N_{s}}\sum_{lk}e^{2\pi iq\left( l-k\right) /N_{s}}\sigma
_{1lk},  \label{mdf}
\end{equation}
$N_{q}$ is the expected total number of particles with momentum $q$ (in
units of $h/N_{s}a$, where $a$ is the lattice spacing). $N_{0}$ may be
identified with the condensate population.

In the\ deep Mott regime ($J=0$), $\sigma _{1lk}=n\delta _{lk}$ and $N_{q}=n$
is the same for all modes. In the opposite limit ($U=0$) $\sigma _{1lk}=n$
for every pair of sites and $N_{q}=N\delta _{q0}.$

Our goal is to obtain analytic expressions for this observable in the
intermediate regime $U/nJ\sim 1,$ with $n\sim 1$ as well.

\subsection{Some approaches to the one-body density matrix}

To motivate our perspective below, let us begin with a brief discussion of
some of the most common approaches to this problem in the literature and
place our work in this context. We feel that other than the few full-fledged
numerical calculations (\cite{mc}\cite{CJ04}), none of the analytic
approaches fully cover the transition regime described above. Moreover, even
if a numerical calculation is feasible it is useful to have a reliable
analytic approach to match against.

To begin with, since our interest is $\sigma _{1}$, approaches based on the
Gutzwiller ansatz or mean field theory \cite{mft} would not be sufficient.
These methods are very powerful to investigate the phase diagram, but
because they treat different sites as independent, they severely distort the
one-body density matrix.

These approaches may be improved on, of course. The Gutzwiller ansatz may be
taken as just the first step in a consistent perturbative expansion \cite
{SMB04}, and the mean field decoupling ansatz may be applied to full cells
rather than individual sites \cite{BV05}. However, the required order in
perturbation theory (or the size of the fundamental cell) to get a reliable
result scales with the size of the lattice, and soon the difficulty becomes
comparable to a full numerical solution.

Starting from the superfluid regime, the simplest way to get $\sigma _{1}$
is the Bogoliubov approach \cite{KPS02}. Since we shall consider the case in
which the gas is at fixed total particle number, rather than fixed chemical
potential, we must consider instead the particle number conserving (PNC)
formalism \cite{conserving}. However, for the purpose of this preliminary
discussion we may make abstraction of the difference.

A simple minded mean field approach, in which we simply replace $a_{j}$ by
its ``expectation value'' $z_{j}$, is bound to fail. Since the BH
Hamiltonian has the global phase invariance $a_{j}\rightarrow e^{i\theta
}a_{j},$ in view of Goldstone theorem the mean field theory must be gapless
\cite{gapless}. In other words, simple- minded mean field theory can only
describe the superfluid phase.

Since the one - body density matrix is the time coincidence limit of the two
- point function $\left\langle a_{j}^{\dagger }\left( t\right) a_{k}\left(
t^{\prime }\right) \right\rangle ,$ one could think of finding equations of
motion for these functions directly, without including a mean field \cite
{phider}, but this approach also fails. In a nutshell, the difficulty is as
follows. The Heisenberg equation of motion for $a_{j}^{\dagger }\left(
t\right) $ is

\begin{equation}
\left( -i\right) \frac{\partial }{\partial t}a_{j}^{\dagger }\left( t\right)
=\left[ H,a_{j}^{\dagger }\left( t\right) \right]
=-\sum_{i}J_{ij}a_{i}^{\dagger }+Ua_{j}^{\dagger 2}a_{j},
\end{equation}
whereby

\begin{equation}
i\frac{\partial }{\partial t}\left\langle a_{j}^{\dagger }\left( t\right)
a_{k}\left( t^{\prime }\right) \right\rangle =\sum_{i}J_{ij}\left\langle
a_{j}^{\dagger }\left( t\right) a_{k}\left( t^{\prime }\right) \right\rangle
-U\left\langle \left( a_{j}^{\dagger 2}a_{j}\right) \left( t\right)
a_{k}\left( t^{\prime }\right) \right\rangle ,
\end{equation}
and we face a closure problem, namely, how to express the four point
function in the last term in terms of two point functions. A typical
resolution is a Hartree - like scheme, where we approximate

\begin{equation}
\left\langle \left( a_{j}^{\dagger 2}a_{j}\right) \left( t\right)
a_{k}\left( t^{\prime }\right) \right\rangle \sim 2\left\langle
a_{j}^{\dagger }\left( t\right) a_{j}\left( t\right) \right\rangle
\left\langle a_{j}^{\dagger }\left( t\right) a_{k}\left( t^{\prime }\right)
\right\rangle \simeq 2n\left\langle a_{j}^{\dagger }\left( t\right)
a_{k}\left( t^{\prime }\right) \right\rangle.
\end{equation}

However, in the weak hopping limit we expect the system will be close to the
MI ground state

\begin{equation}
\left| MI\right\rangle =\prod_{i}\left| n\right\rangle _{i},
\label{mottstate}
\end{equation}
(that is, each site is in a state of well defined occupation number) where
we can compute

\begin{equation}
\left\langle a_{j}^{\dagger }\left( t\right) a_{k}\left( t^{\prime }\right)
\right\rangle \sim n\delta _{ij},
\end{equation}

\begin{equation}
\left\langle \left( a_{j}^{\dagger 2}a_{j}\right) \left( t\right)
a_{k}\left( t^{\prime }\right) \right\rangle \sim n\left( n-1\right) \delta
_{ij}\approx \left( n-1\right) \left\langle a_{j}^{\dagger }\left( t\right)
a_{k}\left( t^{\prime }\right) \right\rangle.
\end{equation}
We see the Hartree approximation is off by a factor of two, even if $n\gg 1$
\cite{LLL03}.

A possible way around this problem is to obtain a formal equation of motion
for an object (say a two point function) for finite $U$ and $J,$ and then
approximate the coefficients in the formal equation (for example, a
self-energy) by their exact value at $J=0$ or for very large $J$, as needed
\cite{GWFMGB05b,SD04}. However, the actual expressions derived in this way
are not reliable at the transition region, which is where our main interest
rests.

In the opposite Mott insulator regime, the most straightforward approach is
Rayleigh - Schrodinger perturbation theory in the parameter $J$ \cite
{FM94,FM96,EM99,am2}. However, the complexity of the calculation increases
steeply with each increasing order, and so its accuracy for finite values of
$J$ is hard to assess. Comparison against exact solutions for $n=1$ and $%
N_{s}=5,$ $7$ and $9$ shows that first order perturbation theory breaks down
before the transition (see below). This is consistent with the expectation
that perturbation theory breaks down when $Jn>U.$

Dilute gases with very strong repulsion may be treated as a free Fermi gas
\cite{tg}. This approach has been recently successfully extended to
densities $n>1$ \cite{PRWC05}.

Returning to our the above failed Hartree attempt, it is clear that the
closure problem arises in the $U$ term because it is \ the nonlinear term,
while the $J$ term is linear. One obvious alternative is to reformulate the
theory in such a way that this situation is reversed. This is accomplished
in the so-called slave boson /slave fermion method \cite{sbm}.

The slave boson method requires the introduction of a large number of
auxiliary fields and new constraints on the theory. In this paper we shall
explore a similar strategy (that is, making the repulsion term linear, the
hopping term nonlinear) while keeping closer to the original fields in the
Hamiltonian.

One possible way to implement this is to observe that the interaction term
is actually quadratic on the site occupation number $n_{i}=a_{i}^{\dagger
}a_{i}$, since $a_{i}^{\dagger 2}a_{i}^{2}=n_{i}\left( n_{i}-1\right) $ .
This suggests to consider as fundamental a ``phase'' variable $\varphi _{i}$
canonically conjugated to the occupation numbers $n_{i}$ \cite{MAD27,HAL81}

\begin{equation}
\left[ n_{j},\varphi _{i}\right] =-i\delta _{ij},
\end{equation}
(here and after we assume $\hbar =1$). The original creation and destruction
operators are

\begin{equation}
a_{i}=\left[ \exp -i\varphi _{i}\right] \sqrt{n_{i}}  \label{mad1}
\end{equation}

\begin{equation}
a_{i}^{\dagger }=\sqrt{n_{i}}\left[ \exp i\varphi _{i}\right].  \label{mad2}
\end{equation}

The implementation of this idea hits some well known difficulties \cite{pvqm}%
. If the operator $\varphi _{i}$ exists and is hermitean, then the operators
$\exp -ir\varphi _{i}$ are unitary and shift the state $\left|
n\right\rangle $ into $\left| n-r\right\rangle $. But such operators
annihilate the vacuum state $\left| 0\right\rangle ,$ so they cannot be
unitary. We shall return to these difficulties below.

In terms of the density and phase variables, the classical Hamiltonian
becomes

\begin{equation}
H=\sum_{i}\;\left\{ -\sum_{j<i}2J_{ij}\sqrt{n_{i}n_{j}}\cos \left[ \varphi
_{i}-\varphi _{j}\right] +\frac{U}{2}n_{i}\left( n_{i}-1\right) \right\}.
\end{equation}

If we further approximate $\sqrt{n_{i}n_{j}}\equiv n$ in the hopping term,
then we obtain the quantum phase model \cite{qpm,SAC99}. This model displays
a phase transition, and it has been used to investigate nonequilibrium
aspects of the Mott transition \cite{DSZB04}.

On closer examination, the approximation involved is valid when $Un>J$ \cite
{PADHL05}. Therefore, for $n\sim 1$ it fails at the transition region. In
conclusion, while the quantum phase model is the best option on the shelf,
it must be generalized to lower densities to be truly reliable in the
relevant regime \cite{BBZ03}.

One possibility is to allow for particle fluctuations, but only as far as
any given site is never more than one particle above or below the average.
Then it is possible to map the problem onto the $XY$ model or else use a
path integral representation in terms of spin 1 coherent states \cite{xym}.
These model also display a phase transition, and a Gross - Pitaievsky
description has been recently developed. However, we are not aware of
attempts to carry the perturbative evaluation of these models to higher
orders. Below we shall explore an alternative strategy with the same overall
goals.

Finally we observe that the so-called truncated Wigner approximation and
other phase space methods have been successfully applied in the $n>>1$ limit
\cite{TWA,JG04}.

From this description we see the lack of suitable treatment in the
literature of the one body density matrix at the transition region for low
densities. Not only there is no single approach which is fully reliable
throughout, but moreover those which are successful on one asymptotic regime
are based on a quite different physical model than the ones which succeed on
the other (compare, e.g., Bogoliubov methods against the Tonks - Girardeau
gas approach or strong coupling perturbation theory). A model which is able
to describe the transition region within a single physical model and keeping
an uniform accuracy would be a definite step forward. This is our aim here.
To be fully understood, however, we must identify some desirable features
any new approach to this problem must possess to be truly useful.

\subsection{Our approach in the context of ongoing research}

As we mentioned above, our interest in this problem of the loading of BEC
atoms on to an optical lattice is motivated by the feasibility of using this
process to initialize a quantum computer. This longer range goal sets
certain constraints on the model which we choose to perform our analysis.

The first consideration is that, although in this paper we shall only
discuss the equilibrium case, in last analysis one needs a full
nonequilibrium formulation of the problem. With this goal in mind, we adopt
the Schwinger - Keldysh or closed time - path (CTP) \cite{phider,ctp}
formalism from scratch. As a side benefit, we shall see below that this
choice is also helpful in overcoming the formal difficulties of the density
- phase representation.

A related requirement is that there should be a well defined way to carry
the perturbative evaluation of the model to any order, but because this will
be unavoidably complex, already the first order in the expansion must give
sensible results. In particular, it is desirable to have the model in path
integral language, as it is most adapt to further implementation of
perturbation theory.

Actually, the simplest quantum phase model formulation fails this test; with
some oversimplification, the problem is that $\sqrt{n+\delta n}\sim \sqrt{n}%
+\delta n/2\sqrt{n}$ is a bad approximation if $\delta n>n$ \cite{AKS01}. We
shall seek a new set of variables in which the perturbative evaluation of $%
\sigma _{1}$ is more accurate than in the original ones. We shall show this
by comparing the first order approximation to our model with the exact
solutions in the case of small systems, and to the PNC and strong coupling
perturbation theories for larger densities.

It is seen in actual experiments that collisions with noncondensed particles
and loss are not significant except on the longest time scales (above $1$s
\cite{FDWWH98}). Therefore we shall consider the case of an isolated gas, i.
e., the total number of particles will be constant \cite{conserving}, as
opposed to the case of a gas interacting with a particle reservoir, whereby
the chemical potential remains constant. However, instead of the PNC
approach, we shall develop a formalism which is more suitable to the path
integral formulation of the model. We shall regard the given value of the
total particle number $N=nN_{s}$ as a constraint on allowed states of the
system, rather than just a dynamical condition. The resulting theory will
amount to an independent quantization of the system; our model and the PNC
one will agree only with respect to the time evolution of observables which
commute with $N.$ Of course, $a_{j}^{\dagger }\left( t\right) a_{i}\left(
t\right) $ is one of these observables; not so the creation or destruction
operators separately. A detailed comparison of the path integral and PNC
approaches is given in Ref. \cite{gaugeinv}.

Let us observe that this procedure is less unusual than it may seem. For
example, in studies of the ground state of the system, it is common to adopt
trial wave functions which preclude site occupations farther from the mean
than a few units (a similar policy is sometimes adopted for the numerical
diagonalization of the Hamiltonian). In practice, this means that the
Hilbert space of allowable states is constrained; the reduction is actually
more drastic than the one postulated here.

A similar procedure has been implemented in the field of high temperature
superconductors near the Mott limit, to enforce such constraints as
excluding double occupancy \cite{LN92}.

From the technical point of view, the advantage of taking the given value of
$N$ as a hard constraint is that in the constrained system, the global phase
invariance of the Bose - Hubbard model (BHM) becomes local in time.
Technically, the model becomes a gauge theory, with the constraint $N$ as
gauge generator \cite{Dirac}. This allows us to take advantage of the
powerful methods of gauge theory quantization (of which we shall only have a
glimpse in this first take of the problem) \cite{PS95}.

When seen under the light of our stated long term goal, a number of
shortcomings of our present work clearly stand up, and it is only fair that
we mention some of them. First, to be sure we have an accurate description
of the transition we should also compute other observables, such as the
particle number fluctuations \cite{OTFYK01} and the dynamic structure factor
\cite{am2,expbragg}. We have only considered a homogeneous lattice, while a
lattice superimposed to a harmonic trap would be more relevant to
applications \cite{trap} (the presence of the trap has a drastic effect on
the transition \cite{trappht}). We have not considered lattice fluctuations
\cite{SOT97} nor finite temperature effects \cite{fintem}. We have
considered a condensate of atoms without internal degrees of freedom, while
of course the internal structure is essential for QIP \cite{AHDL03}.

It is also interesting to observe that some of the quantities we compute in
this paper have been measured, both in one and three dimensional systems
\cite{SMSKE04,SSMKE04}. We will comment briefly on these results in Section
VII; a detailed discussion will be given elsewhere \cite{RHC05}.

In spite of these unachieved goals, the formulation of a fully analytic
theory of the one body density matrix is a necessary first step towards
constructing a realistic theory of the loading process, which we now proceed
to tackle.

\subsection{This paper}

The rest of the paper is organized as follows. Over the next four Sections,
we develop a formal presentation of the model. In Section II, we develop the
CTP path integral representation for expectation values of BEC observables.
In Section III, this representation is translated to the density - phase
representation. In Section IV, we shift to a new set of variables, more
suitable for the further perturbative evaluation of these expectation
values. In Section V we explore the simplest approximation, where the theory
is linearized in the inhomogeneous modes.

In Section VI we apply this machinery to the computation of the one - body
density matrix and the momentum distribution function. In the final Section
VII we show the results of a comparison of our model against both exact
numerical results and other approximated approaches, and conclude with some
final remarks.

In Appendix A we present a brief derivation of the other approximate
approaches discussed in the Results Section, namely, first order
perturbation theory in $J/N$ and the PNC approach to first order in $N^{-1}.$
This results are not new, and are included only to prevent any
misunderstandings due to different notations between this and the original
papers. Appendix B discusses the validity of Eq. (\ref{approxy}) below as an
approximation to Eq. (\ref{yofm}).

\section{The CTP path integral approach to BECs}

In this Section we'll put together the basic formulae for the coherent-state
path integral method \cite{NO98} to compute expectation values of
observables within the causal CTP approach \cite{ctp}.

Before we get down to formulae, let us try and convey the idea of the
approach in simple terms. Let us begin with the problem of computing the
vacuum expectation value $\left\langle \hat{O}\right\rangle _{0}$ of some
observable $O$. One possibility is to add a new term $-J$ $\hat{O}$ to the
Hamiltonian. Let us call $H$ the original Hamiltonian, $H_{J}$ the new
Hamiltonian $H_{J}=H-J$ $\hat{O}$. Then, if we can find the ground state
energy $E_{J}$ of the new hamiltonian, first order perturbation theory
impliest that $\left\langle \hat{O}\right\rangle _{0}=-dE_{J}/dJ$ at $J=0.$
We have translated the problem of computing $\left\langle \hat{O}%
\right\rangle _{0}$ into the problem of computing $E_{J}.$

As it turns out, a surprisingly efficient way of computing $E_{J}$ is by
computing the matrix elements of the euclidean evolution operator \cite
{COL85}

\begin{equation}
e^{W_{e}\left[ J\right] }=\left\langle 0\left| e^{-TH_{J}}\right|
0\right\rangle  \label{euclid}
\end{equation}
where $\left| 0\right\rangle \ $is the vacuum state (we may assume that the
external source $J$ is swithched off adiabatically at infinity, so the
vacuum is unambiguously defined). It turns out that when the euclidean lapse
$T\rightarrow \infty ,$ $W_{e}\left[ J\right] \rightarrow -TE_{J},$ so again
$\left\langle \hat{O}\right\rangle _{0}=T^{-1}dW_{e}\left[ J\right] /dJ$ at $%
J=0.$

In a time-dependent situation, however, an euclidean formulation is not
readily available. One may attempt to make do with the analytical
continuation of Eq. (\ref{euclid}) back to physical time ($T$ inside the
bracket is the time-ordering operator)

\begin{equation}
e^{iW_{m}\left[ J\right] }=\left\langle 0OUT\left| T\left[ e^{-i\int
dt\;H_{J}}\right] \right| 0IN\right\rangle  \label{mink}
\end{equation}
but this is untenable. The ``expectation values'' derived from $W_{m}$ are
generally complex, even if $\hat{O}$ is Hermitian, and they do not evolve
causally in time \cite{CH87}. The problem comes from the fact that they are
no longer expectation values, but rather matrix elements between the
asymptotic vacua $\left| 0IN\right\rangle $ and $\left| 0OUT\right\rangle $,
which are not necessarily equivalent in this time-dependent situation.

The solution found by Schwinger \cite{ctp} was not to include one external
source but two, $J^{1}$ and $J^{2},$ and to define a new generating
functional

\begin{equation}
e^{iW_{CTP}\left[ J\right] }=\left\langle 0IN\left| \tilde{T}\left[ e^{i\int
dt\;H_{J^{2}}}\right] T\left[ e^{-i\int dt\;H_{J^{1}}}\right] \right|
0IN\right\rangle  \label{ctpa}
\end{equation}
where $\tilde{T}$ is the anti time-ordering operator. We may also think of $%
J^{1}$ and $J^{2}$ as a single source defined on a ``closed time-path''
which reaches from $t=0$, say, to the far future (wherein the source takes
the values $J^{1}\left( t\right) $) and then bounces back to $t=0$ (the
source swithching to $J^{2}\left( t\right) $), therein the name of the
method. It is readily shown that differentiation of $W_{CTP}$ yields true
expectation values, which are of course real and evolve causally.

Since both quantum states in Eq. (\ref{ctpa}) are defined at the same
reference time $t=0,$ $W_{CTP}$ is readily generalized to non vacuum
situations \cite{CH88}. Let $\rho _{i}$ be the density matrix describing the
state at $t=t_{i}.$ Then expectation values with respect to $\rho _{i}$ may
be obtained from the CTP generating functional

\begin{equation}
e^{iW}=\mathrm{tr}\left\{ U_{2}^{-1}\left( t_{f},t_{i}\right) U_{1}\left(
t_{f},t_{i}\right) \rho \left( t_{i}\right) \right\}  \label{ctpb}
\end{equation}
where

\begin{equation}
U_{i}\left( t_{f},t_{i}\right) =T\left[
e^{-i\int_{t_{i}}^{t_{f}}dt\;H^{i}\left( t\right) }\right]
\end{equation}

Our problem is to build a generating functional which will allow us to
compute the one - body density matrix. Our starting point will be Eq. (\ref
{ctpb}). For reasons of efficiency, we shall seek a path integral
representation of the trace.

In this and the next two Sections, we shall construct this representation.
In this Section we shall use the well known coherent state representation
\cite{NO98}, putting the emphasis on the implementation of CTP boundary
conditions. Then we shall proceed to rewrite the CTP generating functional
in terms of more suitable variables, to optimize the accuracy of its
perturbative expansion.

\subsection{The coherent state representation}

We shall begin by recalling the usual coherent state path integral
representation of transition amplitudes \cite{NO98}. The CTP boundary
conditions shall be introduced in next Subsection.

For simplicity, let us consider a single one-particle state. There is a
basis made of occupation number eigenstates $\left| n\right\rangle $

\begin{equation}
N\left| n\right\rangle =n\left| n\right\rangle
\end{equation}
In particular, there is the vacuum state $\left| 0\right\rangle .$ These
states are orthonormal and complete

\begin{equation}
\left\langle m\right. \left| n\right\rangle =\delta _{mn}
\end{equation}

\begin{equation}
\sum \left| n\right\rangle \left\langle n\right| =1
\end{equation}
The destruction and creation operators relate states of different occupation
numbers

\begin{equation}
a\left| n\right\rangle =\sqrt{n}\left| n-1\right\rangle ;\qquad a^{\dagger
}\left| n\right\rangle =\sqrt{n+1}\left| n+1\right\rangle
\end{equation}
Therefore

\begin{equation}
a^{\dagger }a=N;\qquad \left[ a,a^{\dagger }\right] =\mathbf{1}
\end{equation}

A coherent state $\left| \bar{a}\right\rangle $ is an eigenstate of the
destruction operator

\begin{equation}
a\left| \bar{a}\right\rangle =\bar{a}\left| \bar{a}\right\rangle
\end{equation}
Adopting the normalization $\left\langle 0\right. \left| \bar{a}%
\right\rangle =1$ we find

\begin{equation}
\left\langle n\right. \left| \bar{a}\right\rangle =\frac{\bar{a}^{n}}{\sqrt{%
n!}}
\end{equation}
Let $\left| \bar{b}\right\rangle $ be a second coherent state; then

\begin{equation}
\left\langle a\right. \left| \bar{b}\right\rangle =\mathrm{exp}\left\{ a^{*}%
\bar{b}\right\}
\end{equation}
The vacuum is the coherent state with $\bar{a}=0$.

While not orthogonal, the coherent states are complete, in the sense that

\begin{equation}
\int \frac{da^{*}da}{2\pi i}\mathrm{exp}\left\{ -a^{*}a\right\} \left|
a\right\rangle \left\langle a\right| =\mathbf{1}
\end{equation}

We may use the completeness relationship to write down the trace of an
operator $A$

\begin{equation}
\mathrm{tr}A=\sum \left\langle n\right| A\left| n\right\rangle =\int \frac{%
da^{*}da}{2\pi i}\mathrm{exp}\left\{ -a^{*}a\right\} \left\langle a\right|
A\left| a\right\rangle
\end{equation}

Now consider the transition amplitude between the state $\left|
a_{i}\right\rangle $ at time $t_{i}=0$ and the state $\left| \bar{a}%
_{f}\right\rangle $ at time $t_{f}$, where $\left| \bar{a}_{f}\right\rangle $
is the eigenstate of the Heisenberg operator $a\left( t_{f}\right) $ with
proper value $\bar{a}.$ Since $a\left( t_{f}\right) =e^{iHt_{f}}ae^{-iHt_{f}}
$, we have $\left( \hbar =1\right) $

\begin{equation}
\left| \bar{a}_{f}\right\rangle =e^{iHt_{f}}\left| \bar{a}\right\rangle
\end{equation}
and

\begin{equation}
\left\langle \bar{a}_{f}\right. \left| a_{i}\right\rangle =\left\langle \bar{%
a}\right| e^{-iHt_{f}}\left| a_{i}\right\rangle
\end{equation}

Let $N$ be some large number and $\varepsilon =t_{f}/N.$ Write $a_{i}=a_{0},$
$\bar{a}=a_{N}.$ Then, inserting $N-1$ identity operators, we have

\begin{equation}
\left\langle \bar{a}_{f}\right. \left| a_{i}\right\rangle =\int \left\{
\prod_{n=1}^{N-1}\frac{da_{n}^{*}da_{n}}{2\pi i}\mathrm{exp}\left\{
-a_{n}^{*}a_{n}\right\} \left\langle a_{n+1}\right| e^{-iH\varepsilon
}\left| a_{n}\right\rangle \right\} \left\langle a_{1}\right|
e^{-iH\varepsilon }\left| a_{0}\right\rangle
\end{equation}
which may be written as (assuming the Hamiltonian $H=H\left( a^{\dagger
},a\right) $ is in normal form)

\begin{equation}
\left\langle \bar{a}_{f}\right. \left| a_{i}\right\rangle =\int \left[
Da\right] _{N-1}\;\mathrm{exp}\left\{ iS_{N}\left[ a^{*},a\right] \right\}
e^{a_{N}^{*}a_{N}}
\end{equation}
where

\begin{equation}
\left[ Da\right] _{N-1}=\prod_{n=1}^{N-1}\frac{da_{n}^{*}da_{n}}{2\pi i}
\end{equation}

\[
S_{N}\left[ a^{*},a\right] =\sum_{n=1}^{N}\left\{ ia_{n}^{*}\left(
a_{n}-a_{n-1}\right) -\varepsilon H\left( a_{n}^{*},a_{n-1}\right) \right\}
\]

Going to the continuum limit, where $a_{n}-a_{n-1}\sim \varepsilon \partial
a/\partial t$, we get

\begin{equation}
\left\langle a\left( t_{f}\right) _{f}\right. \left| a\left( t_{i}\right)
\right\rangle =\int \left[ Da\right] \;\mathrm{exp}\left\{ iS\left[
a^{*},a\right] \right\} e^{a^{*}a\left( t_{f}\right) }
\end{equation}

\begin{equation}
S\left[ a^{*},a\right] =\int dt\;\left\{ ia^{*}\frac{\partial a}{\partial t}%
-H\left( a^{*},a\right) \right\}   \label{limit}
\end{equation}
The integration is over paths where the initial value of $a$ and the final
value of $a^{*}$ are fixed, and given by $a\left( t_{i}\right) $ and $%
a^{*}\left( t_{f}\right) ,$ respectively.

\subsection{The CTP boundary conditions}

We now have all the necessary elements to evaluate the CTP generating
functional Eq. (\ref{ctpb}). The idea is that the initial density matrix $%
\rho $ is propagated forwards in time with some Hamiltonian $H^{1}$ and then
backwards with a Hamiltonian $H^{2}$. Insert three identity operators in Eq.
(\ref{ctpb}) to obtain

\begin{eqnarray}
e^{iW} &=&\int \frac{da_{N}^{*}da_{N}}{2\pi i}\frac{da_{0}^{1*}da_{0}^{1}}{%
2\pi i}\frac{da_{0}^{2*}da_{0}^{2}}{2\pi i}\mathrm{exp}\left\{ -\left(
a_{N}^{*}a_{N}+a_{0}^{1*}a_{0}^{1}+a_{0}^{2*}a_{0}^{2}\right) \right\}
\nonumber \\
&&\left\langle a_{N}\right| U_{2}\left( t_{f},t_{i}\right) \left|
a_{0}^{2}\right\rangle ^{*}\left\langle a_{N}\right| U_{1}\left(
t_{f},t_{i}\right) \left| a_{0}^{1}\right\rangle \left\langle
a_{0}^{1}\right| \rho \left( t_{i}\right) \left| a_{0}^{2}\right\rangle
\end{eqnarray}
Now use the corresponding path integral representations

\begin{eqnarray}
e^{iW} &=&\int \frac{da_{N}^{*}da_{N}}{2\pi i}\frac{da_{0}^{1*}da_{0}^{1}}{%
2\pi i}\frac{da_{0}^{2*}da_{0}^{2}}{2\pi i}\mathrm{exp}\left\{
a_{N}^{*}a_{N}-a_{0}^{1*}a_{0}^{1}-a_{0}^{2*}a_{0}^{2}\right\} \left\langle
a_{0}^{1}\right| \rho \left( t_{i}\right) \left| a_{0}^{2}\right\rangle
\nonumber \\
&&\int \left[ Da^{2}\right] _{N-1}^{*}\;\mathrm{exp}\left\{
-iS_{N}^{2}\left[ a^{2*},a^{2}\right] ^{*}\right\} \int \left[ Da^{1}\right]
_{N-1}\;\mathrm{exp}\left\{ iS_{N}^{1}\left[ a^{1*},a^{1}\right] \right\}
\label{pir1}
\end{eqnarray}
The configuration on the forward branch has $a^{1}\left( 0\right) =a_{0}^{1}$
and $a^{1*}\left( t_{f}\right) =a_{N}^{*}.$ On the backward branch, we have $%
a^{2*}\left( 0\right) =a_{0}^{2*}$ and $a^{2}\left( t_{f}\right) =a_{N}.$
Once $W$ is known, causal expectation values may be computed by
differentiation.

Eq. (\ref{pir1}) is the main result of this Section. In order to make use of
it, however, we must rewrite it in a more suitable set of variables. This
translation is the subject of the next two Sections.

\section{Density and phase variables in the CTP formulation}

In this Section we present the basic elements of the path integral
formulation of a system of bosonic atoms in an optical latice in terms of
number and phase variables, while enforcing a fixed total particle number.
This will set the stage for a further canonical transformation to a more
convenient set of degrees of freedom, to be carried out in the next Section.

\subsection{Madelung representation for the creation and destruction
operators}

Our starting point is the Madelung representation for the creation and
destruction operators, Eqs. (\ref{mad1}) and (\ref{mad2}). The phase
observables $\varphi _{i}$ have eigenstates $\left| \varphi
_{i}\right\rangle ,$ which are a complete basis if $\varphi _{i}$ runs over
a full circle. To account for the periodic nature of these variables, we
define the inner product

\begin{equation}
\left\langle \left\langle \varphi _{i}\mid \varphi _{i}^{\prime
}\right\rangle \right\rangle =\sum_{k}\delta \left( \varphi _{i}-\varphi
_{i}^{\prime }-2\pi k\right) ,  \label{crazyid}
\end{equation}
with $k$ running over all integers. A transition element is decomposed into
transitions between phase eigenstates, mediated by these identity operators.
However, as shown by Kleinert \cite{KLE90}, all but one of these sums may be
avoided if we allow $\varphi _{i}$ to run over all real numbers, and not
just a circle. Since the line is the covering space of the circle, we shall
call this extended theory the covering theory.

In the covering theory, the discrete observable $n_{i}$ is replaced by a
continuous observable $\rho _{i},$ whose eigenstates $\left| \rho
_{i}\right\rangle $ [such that $\left\langle \rho _{i}^{\prime }\left| \rho
_{j}\right. \right\rangle =\delta _{ij}\delta \left( \rho ^{\prime }-\rho
\right) $]) generate the Hilbert space. Then the physical subspace is the
one generated by the $\left| \rho _{i}\right\rangle $ where $\rho _{i}$
happens to be a nonnegative integer.

In the expanded Hilbert space, we have the $\rho $ representation of a state
$\left| \psi \right\rangle $

\begin{equation}
\left\langle \rho _{i}\left| \psi \right. \right\rangle =\psi \left( \rho
_{i}\right),
\end{equation}
In this representation, the operator

\begin{equation}
\varphi _{i}=i\frac{\partial }{\partial \rho _{i}},
\end{equation}
meaning that

\begin{equation}
\left\langle \rho _{i}\left| \varphi _{i}\right| \psi \right\rangle =i\frac{%
\partial }{\partial \rho _{i}}\psi \left( \rho _{i}\right).
\end{equation}
Therefore

\begin{equation}
\left\langle \rho _{i}\left| \exp (-ir\varphi _{i})\right| \psi
\right\rangle =\psi \left( \rho _{i}+r\right).
\end{equation}
So if $\left| \psi \right\rangle =\left| n\right\rangle ,$ $\psi \left( \rho
_{i}\right) =\delta \left( \rho _{i}-n\right) $ and $\exp (-ir\varphi _{i})
\left| n\right\rangle =\left| n-r\right\rangle ,$ as expected.

\begin{equation}
\left\langle \rho _{i}\left| \varphi _{i}\right| \varphi _{i}\right\rangle =i%
\frac{\partial }{\partial \rho _{i}}\left\langle \rho _{i}\mid \varphi
_{i}\right\rangle =\varphi _{i}\left\langle \rho _{i}\mid \varphi
_{i}\right\rangle,
\end{equation}
so

\begin{equation}
\left\langle \rho _{i}\mid \varphi _{i}\right\rangle =e^{-i\varphi _{i}\rho
_{i}}.
\end{equation}

The reason this scheme works better in the CTP formulation than in other
approaches is that it affords us the freedom to choose an arbitrary initial
condition. As long as the initial condition is chosen within the physical
subspace, the dynamics of the system itself warranties that there will be no
unphysical results. In particular, we turn the covering theory into the
physical theory by inserting the one missing identity of the form Eq. (\ref
{crazyid}) into the path integral in a suitable way (see below).

\subsection{CTP path integrals in number and phase variables}

Before considering the BHM, let us discuss how to build CTP path integrals
for the BHM in terms of number and phase variables. The key is to clarify
the boundary conditions the histories within the path integral must satisfy.

Our starting point is the path integral representation Eq. (\ref{pir1}) for
the CTP generating functional. The action is given in Eq. (\ref{limit}),
whereby $ia^{*}$ is formally the momentum conjugate to $a$. To transform the
action to density and phase variables, define a generating functional

\begin{equation}
Q=\frac{-i}{2}a^{2}e^{2i\varphi }
\end{equation}
so

\begin{equation}
ia^{*}=-\frac{\partial Q}{\partial a};\qquad \rho =\frac{\partial Q}{%
\partial \varphi }
\end{equation}
Then

\begin{equation}
ia^{\ast }da-Hdt=\rho d\varphi -Hdt-dQ
\end{equation}
and

\begin{equation}
S=\int dt\;\left\{ \rho \frac{\partial \varphi }{\partial t}-H\left( \rho
,\varphi \right) \right\} -\left[ Q\left( t_{f}\right) -Q\left( t_{i}\right)
\right]  \label{action}
\end{equation}

\begin{equation}
H=-\sum_{ij}J_{ij}\sqrt{\rho _{i}\rho _{j}}\left[ \exp i\left[ \varphi
_{j}-\varphi _{i}\right] \right] +\sum_{i}\frac{U}{2}\rho _{i}\left( \rho
_{i}-1\right)
\end{equation}
where $Q=\left( -i/2\right) N.$

The other factors in the measure may also be expressed in terms of $N.$ In
this representation, we consider paths which begin at values $\varphi
^{1}\left( 0\right) $ and $\varphi ^{2}\left( 0\right) $ and end at a common
phase $\varphi ^{1}\left( T\right) =\varphi ^{2}\left( T\right) =\varphi
_{f} $. Observe that both the initial and final values of the densities are
undetermined.

Of course, the physical phase variable $\varphi $ must be identified with
periodicity $2\pi $. We do this by inserting the identity operator Eq. (\ref
{crazyid}) at some point within the path integral. So we have two kinds of
expectation values: the expectation values of the covering theory (without
the insertion) and the physical expectation values (with the insertion),
which we shall call $\left\langle \left\langle {}\right\rangle \right\rangle
$ and $\left\langle {}\right\rangle ,$ respectively. The relationship
between these constructs will be further clarified below.

\subsection{Enforcing a fixed particle number}

The quantum theory of the BEC may be regarded as the quantization of the
nonrelativistic classical field theory defined by the action functional Eq. (%
\ref{action}) where the canonical variables are $\rho _{i}\left( t\right) $
and their conjugate momentum $\varphi _{i}\left( t\right) .$ We are
interested in the case in which particle number takes on a definite value $N$%
. We may reinforce this point by adding a constraint on the theory. This is
achieved by introducing a Lagrange multiplier $\mu \left( t\right) ,$ and
rewriting the action as

\begin{equation}
S_{fixed}=S+\int dt\;\mu \left( t\right) \sum_{i}\left( \rho _{i}-n\right) \;
\label{newaction}
\end{equation}

The original action Eq. (\ref{action}) is invariant under a global
transformation $\varphi _{i}\left( t\right) \rightarrow \varphi _{i}\left(
t\right) +$ constant but the new action Eq. (\ref{newaction}) is invariant
under the local (in time) transformations

\begin{equation}
\varphi _{i}\left( t\right) \rightarrow \varphi _{i}\left( t\right) +\theta
\left( t\right) ,\qquad \mu \rightarrow \mu -\frac{d\theta }{dt}
\label{local}
\end{equation}
When $\theta $ is infinitesimal, these are just canonical transformations
generated by the constraint. Therefore it must be quantized using the
methods developed for gauge theories, such as the Fadeev-Popov method.

This comes about because now the path integral is redundant, since we may
transform the fields as in Eq. (\ref{local}). We may fix the redundancy by
factoring out the gauge group. Choose some function $f_{\theta }=f\left[ \mu
_{\theta },\varphi _{i\theta }\right] ,$ such that $df_{\theta }/d\theta
\neq 0.$ Then

\begin{equation}
e^{iW}=\Theta \int D\rho _{i}^{a}\left( t\right) D\varphi _{i}^{a}\left(
t\right) D\mu ^{a}\;e^{i\left[ S_{tot}^{1}-S_{tot}^{2}\right] }Det\left[
\frac{\delta f_{\theta }}{\delta \theta }\right] _{\theta =0}  \label{gf2}
\end{equation}
where

\begin{equation}
\Theta =\int D\theta
\end{equation}
is the volume of the gauge group we wish to factor out,

\begin{equation}
S_{tot}^{a}=S^{a}+\int dt\;\sum_{i}\mu \left( t\right) \left[ \rho
_{i}-n\right] +\frac{1}{2s}\int dt\;f_{0}^{2}  \label{gaugefixed}
\end{equation}
and $s$ is the ``gauge fixing parameter'', which may be chosen freely. The
determinant may be expressed as a path integral over Grassmann ``ghost''
fields \cite{PS95}. For simplicity, we shall adopt a gauge fixing condition $%
f$ which transforms linearly,

\begin{equation}
f=\frac{d\mu }{dt}
\end{equation}
so that its determinant is a constant and may be ignored.

In the path integral, $\mu ^{1}\left( 0\right) $ and $\mu ^{2}\left(
0\right) $ are integrated over. In principle, there are no restrictions at $%
T,$ but we may assume $\mu ^{1}\left( T\right) =\mu ^{2}\left( T\right) $
with no loss of generality. Physically, $\mu $ has the meaning of a
fluctuating chemical potential.

We note that the freedom to choose the gauge fixing condition $f$ and the
gauge fixing parameter $s$ is the key to the power of the method. In this
paper, we shall restrict ourselves to the simple choice above for $f$ and to
the ``Landau'' gauge $s\rightarrow 0$. Other choices may be used to meet the
demands of more advanced applications or to optimize the convergence of
perturbation theory.

A different strategy to introduce freely chosen functions in the formalism
and then exploit the freedom therefrom is the so-called stochastic gauge
method \cite{DDK04}.

Generating functionals as defined in Eqs. (\ref{pir1}) and (\ref{gf2}),
after adding an external source coupled to an operator $\hat{O},$ may be
used to generate correlation functions involving this operator. If $\hat{O}$
commutes with $N$ both representations are equivalent. Otherwise, they will
yield different results. Fortunately, when we compute the one body density
matrix we are within the domain of equivalence of both formalisms.

\subsection{Vanishing of the order parameter}

As a check on the formalism being developped, let us verify that the order
parameter $\left\langle a_{i}\left( t\right) \right\rangle $ vanishes
identically. This must hold in any system with a finite number of particles.

To compute the order parameter, let us add a new source coupled to $\varphi
_{i}^{1}$

\begin{equation}
\int dt^{\prime }\;\sum_{j}j_{j}^{1}\left( t^{\prime }\right) \varphi
_{j}^{1}\left( t^{\prime }\right)
\end{equation}
Now observe that

\begin{equation}
\left\langle a_{k}^{\dagger }\left( t^{\prime }\right) \right\rangle
=\left\langle \sqrt{\rho _{k}}\left( t^{\prime }\right) \right\rangle _{\bar{%
j}_{1}}
\end{equation}
where

\begin{equation}
\left\langle \sqrt{\rho _{k}}\left( t\right) \right\rangle _{\bar{j}%
_{k}}=\int D\rho _{i}^{a}\left( t\right) D\varphi _{i}^{a}\left( t\right)
D\mu ^{a}\;\sqrt{\rho _{k}^{1}}\left( t\right) \exp \left\{i\left[
S_{tot}^{1}-S_{tot}^{2}+\int dt^{\prime }\;\sum_{j}\bar{j}_{kj}\left(
t^{\prime }\right) \varphi _{j}^{1}\left( t^{\prime }\right) \right] \right\}
\end{equation}
and

\begin{equation}
\bar{j}_{ki}\left( t\right) =-\delta \left( t-t_{1}\right) \delta _{ik}
\end{equation}

Let us now show that this vanishes. Make the change of variables within the
path integral

\begin{equation}
\varphi _{i}^{a}\left( t\right) =\bar{\varphi}^{a}\left( t\right) +\delta
\varphi _{i}^{a}\left( t\right) ,\qquad \sum_{i}^{a}\delta \varphi
_{i}^{a}\left( t\right) =0
\end{equation}
The homogeneous phases $\bar{\varphi}^{a}\left( t\right) $ appear linearly
in the action. When we integrate them out, they enforce the constraints

\begin{equation}
\sum_{i}\left[ \frac{d\rho _{i}^{1}}{dt}-\bar{j}_{ki}\right] =0;\qquad
\sum_{i}\frac{d\rho _{i}^{2}}{dt}=0
\end{equation}
But these constraints are impossible to meet, since they contradict the
further constraints from the integration over $\mu $ in the $s\rightarrow 0$
gauge.

\section{A new set of degrees of freedom}

In this Section, we perform a canonical transformation from the phase and
density variables introduced above, to a new set of degrees of freedom which
are more adept for the perturbative evaluation of the one - body density
matrix $\left\langle a_{l}^{\dagger }\left( t_{1}\right) a_{k}\left(
t_{1}\right) \right\rangle .$

As in the evaluation of the order parameter in the previous Section, the
basic idea is to write the creation and destruction operators in their polar
representation, and then consider the exponential terms as the result of
external sources coupled to the phases.

In the general case, a closed evaluation of the one-body density matrix is
not possible. We may try assuming that all quantities may be decomposed into
an homogeneous component and a small inhomogeneous part

\begin{equation}
\rho _{i}^{1}=\bar{\rho}^{1}+\delta \rho _{i}^{1}
\end{equation}
etc. However, one is concerned about the $\sqrt{\rho _{i}}$ factors in the
expectation value. Concretely, while the action becomes quadratic in the $%
J\rightarrow 0$ limit, and so a linearized approximation would seem
reasonable at small enough hopping, a term like $\sqrt{\rho _{i}}$ does not
become Gaussian in any controlled way.

Our proposal is to introduce a new set of canonically conjugated variables,
so that no square roots appear in the definition of the one - body density
matrix or the Hamiltonian. This will make the perturbative expansion
starting from a quadratic approximation to the Hamiltonian more
straightforward.

However, as stated in the Introduction, it is not enough to have a well
defined perturbative expansion, but already the first terms must give
sensible results. In our case, the first term in the expansion corresponds
to keeping only the quadratic terms in the Hamiltonian. This simplified
model will be investigated in next Section.

\subsection{A new set of variables}

To avoid the nonanalytic square roots in the creation and destruction
operators, we proceed as follows. In the first branch, we define a new
(complex) variable $\xi _{i}^{1}$ from

\begin{equation}
a_{i}^{1}=\left[ \exp -i\varphi _{i}^{1}\right] \sqrt{\rho _{i}^{1}}=\exp
\left[ -i\xi _{i}^{1}\right]
\end{equation}

\begin{equation}
a_{i}^{1\dagger }=\sqrt{\rho _{i}^{1}}\left[ \exp i\varphi _{i}^{1}\right]
=\rho _{i}^{1}\;\exp \left[ i\xi _{i}^{1}\right]
\end{equation}
This is actually a canonical transformation, since

\begin{equation}
\rho _{i}^{1}=-\frac{\partial Q^{1}}{\partial \varphi _{i}^{1}};\qquad
Q^{1}=\left( \frac{-i}{2}\right) \sum_{i}\left[ \exp -2i\left( \xi
_{i}^{1}-\varphi _{i}^{1}\right) \right] =\frac{-i}{2}\sum_{i}\rho _{i}^{1}
\end{equation}
The new conjugated momentum is again $\rho _{i}^{1}.$ It follows that on the
first branch

\begin{equation}
S^{1}=\int dt\;\left\{ \sum_{i}\rho _{i}^{1}\frac{\partial \xi _{i}^{1}}{%
\partial t}-H\left( \rho ^{1},\xi ^{1}\right) \right\} +\frac{i}{2}\left[
\sum_{i}\rho _{i}^{1}\right] _{0}^{T}
\end{equation}

On the second branch we write instead

\begin{equation}
a_{i}^{2\dagger }=\sqrt{\rho _{i}^{2}}\left[ \exp i\varphi _{i}^{2}\right]
=\exp i\xi _{i}^{2*}
\end{equation}
Again there is a generating function

\begin{equation}
Q^{2}=\frac{i}{2}\sum_{i}\left[ \exp 2i\left( \xi _{i}^{2\ast }-\varphi
_{i}^{2}\right) \right] =\frac{i}{2}\sum_{i}\rho _{i}^{2}
\end{equation}
so $\rho _{i}^{2}$ is the momentum conjugated to $\xi _{i}^{2\ast }.$ The
action

\begin{equation}
S^{2}=\int dt\;\left\{ \sum_{i}\rho _{i}^{2}\frac{\partial \xi _{i}^{2*}}{%
\partial t}-H\left( \rho ^{2},\xi ^{2*}\right) \right\} -\frac{i}{2}\left[
\sum_{i}\rho _{i}^{2}\right] _{0}^{T}
\end{equation}
In the second branch, therefore

\begin{equation}
a_{i}^{2}=\left[ \exp -i\xi _{i}^{2*}\right] \;\rho _{i}^{2}
\end{equation}
Explicitly, the Hamiltonians read

\begin{equation}
H\left( \rho ^{1},\xi ^{1}\right) =-\sum_{ij}J_{ij}\rho _{j}^{1}\left[ \exp
i\left[ \xi _{j}^{1}-\xi _{i}^{1}\right] \right] +\sum_{i}\frac{U}{2}\rho
_{i}^{1}\left( \rho _{i}^{1}-1\right)
\end{equation}

\begin{equation}
H\left( \rho ^{2},\xi ^{2*}\right) =-\sum_{ij}J_{ij}\left[ \exp i\left[ \xi
_{j}^{2*}-\xi _{i}^{2*}\right] \right] \rho _{i}^{2}+\sum_{i}\frac{U}{2}\rho
_{i}^{2}\left( \rho _{i}^{2}-1\right)
\end{equation}
plus the gauge terms, in both cases. Observe that in the new variables, the
action is explicitly analytical.

\subsubsection{Canonical matters}

If we regard the $a_{i}$ as q-numbers with equal time commutators

\begin{equation}
\left[ a_{i},a_{j}^{\dagger }\right] =\delta _{ij}
\end{equation}
we conclude

\begin{equation}
\left[ \rho _{i},\rho _{j}\right] =0
\end{equation}

\begin{equation}
\left[ \rho _{i},a_{j}\right] =-a_{i}\delta _{ij}
\end{equation}
So

\begin{equation}
\left[ \rho _{i},e^{-i\xi _{j}}\right] =-e^{-i\xi _{i}}\delta _{ij}
\end{equation}

Now observe that

\begin{equation}
e^{-i\xi _{j}}\rho _{i}e^{i\xi _{j}}=\rho _{i}+i\left[ \rho _{i},\xi
_{j}\right] +...
\end{equation}
but also

\begin{equation}
e^{-i\xi _{j}}\rho _{i}e^{i\xi _{j}}=\rho _{i}-\left[ \rho _{i},e^{-i\xi
_{j}}\right] e^{i\xi _{j}}
\end{equation}
Therefore

\begin{equation}
\left[ \rho _{i},\xi _{j}\right] =-i\delta _{ij}  \label{cancom}
\end{equation}
Finally, the $\xi _{j}$ commute among themselves.

As a curiosity, we can actually solve for the $\xi _{j}$ operators. The
commutation rule suggests $\xi _{j}=\varphi _{j}+ig\left( \rho _{i}\right) $
. We now have

\begin{eqnarray}
\left\langle \rho _{2}\left| a\right| \rho _{1}\right\rangle &=&\sqrt{\rho
_{1}}\delta _{\rho _{2}+1,\rho _{1}}=\left\langle \rho _{2}\left| e^{-i\xi
}\right| \rho _{1}\right\rangle  \nonumber \\
&=&\left\langle \rho _{2}\left| \left[ e^{\left[ -i\varphi +g\left( \rho
\right) \right] /N}\right] ^{N}\right| \rho _{1}\right\rangle  \nonumber \\
&=&\int D\varphi D\rho \;\exp \left\{ i\int_{0}^{1}dt\;\left[ -\varphi
\left( \frac{d\rho }{dt}+1\right) -ig\left( \rho \right) \right] \right\}
\end{eqnarray}
where we integrate over paths with $\rho \left( 0\right) =\rho _{1}$, $\rho
\left( 1\right) =\rho _{2}.$ From the integration over $\varphi ,$ we get $%
d\rho /dt\equiv -1,$ and so the functional integral vanishes unless $\rho
_{2}=\rho _{1}-1,$ as expected. Finally

\begin{equation}
\sqrt{\rho _{1}}=e^{\int_{\rho _{1}-1}^{\rho _{1}}d\rho \;g\left( \rho
\right) }
\end{equation}
so, if $G\left( \rho \right) $ is the primitive of $g\left( \rho \right) ,$

\begin{equation}
G\left( \rho \right) =G\left( \rho -1\right) +\frac{1}{2}\ln \left[ \rho
\right]
\end{equation}
so

\begin{equation}
G\left( \rho \right) =\frac{1}{2}\ln \Gamma \left[ \rho +1\right]
\end{equation}
and

\begin{equation}
g\left( \rho \right) =\frac{1}{2}\psi \left[ \rho +1\right]
\end{equation}
If $\rho \gg 1,$ we have the Stirling approximation $\ln \Gamma \left[
x\right] \sim \left( x-1\right) \ln \left( x-1\right) -\left( x-1\right) $
and so $\psi \left[ x\right] \sim \ln \left( x-1\right) ,$ as expected. If
we reinstate the $\hbar $ factors, we find we must replace $\rho $ by $\rho
/\hbar ,$ so in the semiclassical limit we always have $\rho \gg 1.$

Let us close this Section with a word on the path of integration and measure
appropriate to the path integral representation. Since the $\left( \xi
_{i},\rho _{i}\right) $ are canonically conjugated variables, the measure of
integration is the Liouville measure at each time slice in the path
integral, $d\xi _{i}\wedge d\rho _{i}.$ To reduce this to a more familiar
form, we observe that since $g\left( \rho _{i}\right) =\left( -i\right)
\left( \xi _{i}-\xi _{i}^{\ast }\right) /2,$ then $d\xi _{i}\wedge d\rho
_{i}=i\;d\xi _{i}\wedge d\xi _{i}^{\ast }/2g^{\prime }\left( \rho
_{i}\right) .$ Therefore at each time slice we must integrate over the whole
complex $\xi _{i}$ plane; if we adopt the noncanonical (but more usual) $%
\left( \xi _{i},\xi _{i}^{\ast }\right) $ pair as independent variables,
then a non trivial measure arises.

This subtlety will not be an obstacle in what follows, since the relevant
expectation values will be computed directly from symmetry arguments or by
using the properties of the Heisenberg ($q$-number) operators involved,
rather than by an explicit evaluation of the path integral.

\subsection{The dynamics in the new variables}

At this point we introduce the eigenvectors $f_{pj}$ of the matrix $J_{ij}$
and the corresponding eigenvalues $j_{p}$. For example, consider the case in
$d=1$, $N_{s}=2K+1.$
\begin{equation}
J_{ij}=J\left[ \delta _{i,j+1}+\delta _{i,j-1}\right]
\end{equation}
The eigenvectors of $J_{ij}$ are

\begin{equation}
f_{pj}=\frac{1}{\sqrt{N_{s}}}\exp \left[ \frac{2\pi ipj}{N_{s}}\right]
;\qquad -K\leq p\leq K
\end{equation}
and the eigenvalues

\begin{equation}
j_{p}=2J\cos \left[ \frac{2\pi p}{N_{s}}\right]
\end{equation}

We now split all variables into a homogeneous and an inhomogeneous part

\begin{equation}
\rho _{i}^{a}\left( t\right) =\rho _{0}^{a}\left( t\right) +r_{i}^{a}\left(
t\right)  \label{decomp}
\end{equation}

\begin{equation}
r_{i}^{a}\left( t\right) =\sum_{p\neq 0}r_{p}^{a}\left( t\right) f_{pi}
\label{rp}
\end{equation}
and similarly

\begin{equation}
\xi _{i}^{a}\left( t\right) =\xi _{0}^{a}\left( t\right) +X_{i}^{a}\left(
t\right)
\end{equation}

\begin{equation}
X_{i}^{a}\left( t\right) =\sum_{p\neq 0}X_{p}^{a}\left( t\right) f_{pi}
\label{xp}
\end{equation}
The Hamiltonian becomes

\begin{eqnarray}
H\left( \rho _{0}^{1},r_{p}^{1},X_{p}^{1}\right) &=&-\rho
_{0}^{1}\sum_{ij}J_{ij}\left[ \exp i\left[ X_{j}^{1}-X_{i}^{1}\right]
\right] -\sum_{ij}J_{ij}r_{j}^{1}\left[ \exp i\left[
X_{j}^{1}-X_{i}^{1}\right] \right]  \nonumber \\
&&+N_{s}\frac{U}{2}\rho _{0}^{1}\left( \rho _{0}^{1}-1\right) +\sum_{i}\frac{%
U}{2}\left( r_{i}^{1}\right) ^{2}  \label{Hamiltonian}
\end{eqnarray}

\section{The linearized approximation}

We see from Eq.(\ref{Hamiltonian}) that the model ressembles the dynamics of
a solid, with the $X_{i}$ playing the role of the ion positions and a
periodic interaction potential between ions. There is a sophisticated
technology to deal with such systems \cite{PADHL05}. The simplest possible
approach is the linearized approximation in which the excitations of the
``solid'' are described as a free phonon gas. We shall now develop the
implications of this view.

In this Section, we shall derive concrete expressions for the Heisenberg
operators corresponding to the $X_{p}$ and $r_{p}$ operators introduced
above, Eqs. (\ref{rp}) and (\ref{xp}). These expressions shall be used in
the next Section to derive an analytic approximation for the one body
density matrix.

\subsection{The lowest order equilibrium theory}

We obtain the lowest order theory by keeping only the quadratic terms in the
classical action. It is a theory of linear fields, and so we may either
attempt to solve the equations of motion for the propagators, or else solve
the Heisenberg equations, which are the same as the classical equations of
motion, and compute the propagators afterwards. We shall adopt the second
path.

In this Section, therefore, we shall work directly in terms of $q$-number
operators, the Heisenberg equations and canonical commutation relations,
rather than from the path integral.

The ``free'' quadratic part of the Hamiltonian is

\begin{eqnarray}
H\left( \rho _{0},r_{p},X_{p}\right) &=&\sum_{i}\frac{U}{2}\left(
r_{i}\right) ^{2}+\frac{\rho _{0}}{2}\sum_{ij}J_{ij}\left[
X_{j}-X_{i}\right] ^{2}-i\sum_{ij}J_{ij}r_{j}\left[ X_{j}-X_{i}\right]
\nonumber \\
&&+N_{s}\frac{U}{2}\rho _{0}\left( \rho _{0}-1\right)
\end{eqnarray}
Call

\begin{equation}
\nu _{p}=2\left( 2J-j_{p}\right) =8J\sin ^{2}\frac{\pi p}{N_{s}}
\end{equation}

\begin{equation}
\sum_{ij}f_{j-p}J_{ij}\left[ X_{j}-X_{i}\right] =\frac{\nu _{p}}{2}X_{p}
\label{id4}
\end{equation}

\begin{equation}
\sum_{ij}J_{ij}\left[ X_{j}-X_{i}\right] ^{2}=2\sum_{ij}J_{ij}\left[ \left(
X_{j}\right) ^{2}-X_{j}X_{i}\right] =\sum_{p}\nu _{p}X_{-p}X_{p}
\end{equation}
so (assuming for $\rho _{0}$ the constrained value $\rho _{0}=n$)

\begin{equation}
H\left( r_{p},X_{p}\right) =\frac{U}{2}\sum_{p}r_{-p}r_{p}+\frac{n}{2}%
\sum_{p}\nu _{p}X_{-p}X_{p}-\frac{i}{2}\sum_{p}\nu _{p}r_{-p}X_{p}
\end{equation}

The Heisenberg equations of motion are just the classical Hamilton equations

\begin{equation}
\frac{dX_{p}}{dt}=Ur_{p}-\frac{i}{2}\nu _{p}X_{p}
\end{equation}

\begin{equation}
\frac{dr_{p}}{dt}=\frac{i}{2}\nu _{p}r_{p}-n\nu _{p}X_{p}
\end{equation}
We seek a solution of the form

\begin{equation}
X_{p}\left( t\right) =A_{p}e^{-i\omega _{p}t}+B_{-p}^{\dagger }e^{i\omega
_{p}t}
\end{equation}
(recall that the $X_{i}$ are not hermitean, but they satisfy $\left(
X^{*}\right) _{p}=\left( X_{-p}\right) ^{*}$)

\begin{equation}
r_{p}\left( t\right) =\left( -i\right) \left[ C_{p}e^{-i\omega
_{p}t}-C_{-p}^{\dagger }e^{i\omega _{p}t}\right]
\end{equation}
therefore

\begin{equation}
A_{p}=\frac{U}{\omega _{p}-\frac{\nu _{p}}{2}}C_{p}
\end{equation}

\begin{equation}
B_{p}=\frac{U}{\omega _{p}+\frac{\nu _{p}}{2}}C_{p}
\end{equation}
with a dispersion relation \cite{HAL81}

\begin{equation}
\omega _{p}=\sqrt{\nu _{p}\left( Un+\frac{\nu _{p}}{4}\right) }
\label{phonspec}
\end{equation}

We now write down the canonical commutation relations (cfr. Eq. (\ref{cancom}%
))

\begin{equation}
\left[ r_{i},X_{j}\right] =\left( -i\right) \left[ \delta _{ij}-\frac{1}{%
N_{s}}\right]
\end{equation}
whereby

\begin{equation}
\left[ r_{p},X_{q}\right] =\left( -i\right) \delta _{p+q,0}
\end{equation}

Substituting our solution of the Heisenberg equations, we obtain $\left[
C_{p},C_{q}\right] =0$ and

\begin{equation}
\frac{2\omega _{p}}{n\nu _{p}}\left[ C_{p},C_{q}^{\dagger }\right] =\delta
_{p,q}
\end{equation}
Therefore

\begin{equation}
C_{p}=\sqrt{\frac{n\nu _{p}}{2\omega _{p}}}\alpha _{p}
\end{equation}
where $\alpha _{p}$ is a canonical destruction operator. The final formulae
read

\begin{equation}
r_{p}\left( t\right) =\left( -i\right) \sqrt{\frac{n\nu _{p}}{2\omega _{p}}}%
\left[ \alpha _{p}e^{-i\omega _{p}t}-\alpha _{-p}^{\dagger }e^{i\omega
_{p}t}\right]
\end{equation}

\begin{equation}
X_{p}\left( t\right) =\frac{1}{\sqrt{2n\nu _{p}\omega _{p}}}\left\{ \left[
\omega _{p}+\frac{\nu _{p}}{2}\right] \alpha _{p}e^{-i\omega _{p}t}+\left[
\omega _{p}-\frac{\nu _{p}}{2}\right] \alpha _{-p}^{\dagger }e^{i\omega
_{p}t}\right\}
\end{equation}

With the same argument, we get $\left( X^{*}\right) _{p}=$

\begin{equation}
\left( X^{*}\right) _{p}=\frac{1}{\sqrt{2n\nu _{p}\omega _{p}}}\left\{
\left[ \omega _{p}-\frac{\nu _{p}}{2}\right] \alpha _{p}e^{-i\omega
_{p}t}+\left[ \omega _{p}+\frac{\nu _{p}}{2}\right] \alpha _{-p}^{\dagger
}e^{i\omega _{p}t}\right\}
\end{equation}

\subsection{Wick theorem and boundary conditions}

Because we have been able to keep the discussion so far at the level of the
Hamiltonian and canonical operators, we did not have to deal with the issue
of the periodicity conditions on the phase variables. However, in order to
actually use this theory to compute physical expectation values, we must
confront this issue.

Let us first ignore all periodicity conditions. This means that we are
concerned with the expectation values $\left\langle \left\langle
{}\right\rangle \right\rangle $ of the covering theory. Because the lowest
order approximation to the Hamiltonian is Gaussian, Wick's theorem implies

\begin{equation}
\left\langle \left\langle e^{iA}\right\rangle \right\rangle
=e^{-\left\langle \left\langle A^{2}\right\rangle \right\rangle /2}
\end{equation}
for any linear function $A$ of the $X_{p}$'s and $r_{p}$'s.

Now we turn to the issue of the expectation values in the physical theory,
where the periodicity conditions are enforced. To this effect it is
convenient to think in terms of the path integral representation of the
expectation values.

One way to implement the boundary conditions is not to simply match the
history at the second branch to the initial state at time $t^{\left(
2\right) }=0,$ but to all possible translations of this initial state,
namely, translating all the phases by an amount $\varphi _{i}\rightarrow
\varphi _{i}+2\pi k_{i},$ for all possible integer $k_{i}.$ Since the
translation operator for phases is $\rho _{i},$ this amounts to defining a
new expectation value
\begin{equation}
\left\langle e^{iA}\right\rangle =C\sum_{k_{i}}\left\langle \left\langle
e^{2\pi i\sum_{i}k_{i}\rho _{i}^{\left( 2\right) }\left( 0\right)
}e^{iA}\right\rangle \right\rangle.
\end{equation}

In this equation, the expectation value on the left hand side is physical,
and the one on the right hand side belongs to the covering theory. Observe
that because we are using a CTP representation of the path integrals, there
are no ordering ambiguities; this is one of the side benefits of the CTP
formulation, even when discussing equilibrium properties.

Adding the new exponentials is equivalent to forcing $\rho _{i}^{\left(
2\right) }\left( 0\right) $ to be an integer. Since the homogeneous value $%
\rho _{0}^{\left( 2\right) }$ is already constrained to be an integer by the
integration over the Lagrange multiplier $\mu ^{\left( 2\right) }\left(
0\right) ,$ it may be ignored. So we shall adopt the definition

\begin{equation}
\left\langle e^{iA}\right\rangle =C\sum_{k_{i}}\left\langle \left\langle
e^{2\pi i\sum_{i}k_{i}r_{i}^{\left( 2\right) }\left( 0\right)
}e^{iA}\right\rangle \right\rangle
\end{equation}

The constant $C$ is defined by the condition that $\left\langle
1\right\rangle =1,$ so

\begin{equation}
C\sum_{k_{i}}\exp \left\{ -2\pi ^{2}\sum_{lm}k_{l}M_{lm}k_{m}\right\} =1
\end{equation}
where

\begin{equation}
M_{lm}=\left\langle \left\langle r_{l}r_{m}\right\rangle \right\rangle
\end{equation}
We now have

\begin{equation}
\left\langle e^{iA}\right\rangle =C\sum_{k_{i}}e^{\frac{-1}{2}\left\langle
\left\langle \left[ A+2\pi \sum_{i}k_{i}r_{i}^{\left( 2\right) }\left(
0\right) \right] ^{2}\right\rangle \right\rangle }  \label{key1}
\end{equation}
The exponent reads

\begin{equation}
=\left\langle \left\langle A^{2}\right\rangle \right\rangle +4\pi
\sum_{i}k_{i}\left\langle \left\langle r_{i}^{\left( 2\right) }\left(
0\right) A\right\rangle \right\rangle +\left( 2\pi \right)
^{2}\sum_{lm}k_{l}M_{lm}k_{m}  \label{key2}
\end{equation}
Finally, observe that

\begin{equation}
M_{lm}=\frac{n}{N_{s}}\sum_{p\neq 0}e^{2\pi ip\left( l-m\right) /N_{s}}\frac{%
\nu _{p}}{2\omega _{p}}
\end{equation}
Since the sum is restricted to nonvanishing momenta, $M_{lm}$ has a zero
mode, corresponding to homogeneous configurations. In the orthogonal
subspace, $M_{lm}$ has an inverse

\begin{equation}
M_{lm}^{-1}=\frac{1}{nN_{s}}\sum_{p\neq 0}e^{2\pi ip\left( l-m\right) /N_{s}}%
\frac{2\omega _{p}}{\nu _{p}}
\end{equation}

Let us check the asymptotic form of these expressions

\subsubsection{The SF limit}

The matrix $M_{lm}$ represents the particle number fluctuation correlations
between sites $l$ and $m.$ In the SF limit, we expect $M_{lm}=n\left[ \delta
_{lm}-\frac{1}{N_{s}}\right] $. Indeed, in this limit $2\omega
_{p}\rightarrow \nu _{p}.$ So

\begin{equation}
\left\langle e^{iA}\right\rangle \sim Ce^{\frac{-1}{2}\left\langle
\left\langle A^{2}\right\rangle \right\rangle }\sum_{k_{i}}e^{-2\pi \left[
\sum_{i}k_{i}\left\langle \left\langle r_{i}^{\left( 2\right) }\left(
0\right) A\right\rangle \right\rangle +\pi nk_{i}^{2}\right] }
\end{equation}
The sum will be dominated by the $k_{i}=0$ term, so

\begin{equation}
\left\langle e^{iA}\right\rangle \rightarrow e^{\frac{-1}{2}\left\langle
\left\langle A^{2}\right\rangle \right\rangle }
\end{equation}

In other words, treating the density variables as continuous (thereby
ignoring periodicity conditions) is not a serious mistake. This is
consistent with the large number fluctuations in this regime.

\subsubsection{The MI limit}

For the same reasons, in the MI limit we expect $M_{lm}\rightarrow 0.$ Then
many terms contribute to the sum, and we may replace the discrete sum by an
integral. Under this approximation, in fact, we are forcing the $r_{i}$ not
just to be an integer, but to vanish.

Completing the square, we get

\begin{equation}
\left\langle e^{iA}\right\rangle =e^{\frac{-1}{2}\left\langle \left\langle
A^{2}\right\rangle \right\rangle }\;\exp \left\{ \frac{1}{2}%
\sum_{jh}\left\langle \left\langle r_{j}^{\left( 2\right) }\left( 0\right)
A\right\rangle \right\rangle M_{jh}^{-1}\left\langle \left\langle
r_{h}^{\left( 2\right) }\left( 0\right) A\right\rangle \right\rangle \right\}
\end{equation}
As expected, in this limit

\begin{equation}
\left\langle e^{i\sum_{k}\beta _{k}r_{k}\left( 0\right) }\right\rangle =1
\label{lemma}
\end{equation}
We can prove this by observing that, for $A=\beta r_{k},$ $\left\langle
\left\langle r_{h}^{\left( 2\right) }\left( 0\right) A\right\rangle
\right\rangle =\beta M_{hk}$. This implies that the particle number
fluctuations $\left\langle r_{k}\left( 0\right) r_{l}\left( 0\right)
\right\rangle $ vanish in this limit.

\section{The one-body density matrix}

We may now turn to computing the one-body density matrix Eq. (\ref
{onebodydenmat})

\begin{equation}
\sigma _{1lk}=\left\langle a_{l}^{\dagger }\left( t_{1}\right) a_{k}\left(
t_{1}\right) \right\rangle \equiv \left\langle \exp i\left[ \xi
_{l}^{2*}-\xi _{k}^{1}\right] \left( t_{1}\right) \right\rangle
\end{equation}
Observe that in our variables, the observable to be computed is a pure
exponential: there are no square roots to be developed. This is the whole
point of introducing the new variables.

We shall use the machinery introduced above. The desired expectation value
is given by Eqs. (\ref{key1}) and (\ref{key2})

\subsection{The one body density matrix in the covering theory}

The first step is to compute the expectation value disregarding periodicity
conditions, that is, extending the path integral to the covering space

\begin{equation}
\left\langle \left\langle a_{l}^{\dagger }\left( t_{1}\right) a_{k}\left(
t_{1}\right) \right\rangle \right\rangle \equiv \left\langle \left\langle
\exp i\left[ \xi _{l}^{2*}-\xi _{k}^{1}\right] \left( t_{1}\right)
\right\rangle \right\rangle
\end{equation}

To compute this expression, one would aim to split the $\xi $ variables into
homogeneous and inhomogeneous parts, and to use the formulae above. However,
there is a difficulty. We have solved for the Heisenberg operators
corresponding to the inhomogeneous parts, but not for those of the
homogeneous terms. As it happens, however, using a symmetry argument saves
this effort. Because of the symmetry and the total particle number
constraint we must have

\begin{equation}
\sigma _{100}=\left\langle a_{k}^{\dagger }\left( t_{1}\right) a_{k}\left(
t_{1}\right) \right\rangle =n  \label{symmetry}
\end{equation}
Using this property we can avoid computing explicitly the expectation values
for the homogeneous operators. Notice however that Eq. (\ref{symmetry})
involves a physical expectation value (as oppossed to an expectation value
within the covering theory).

To return to our main argument, we shall seek an expression not for the
expectation value $\left\langle \left\langle a_{l}^{\dagger }\left(
t_{1}\right) a_{k}\left( t_{1}\right) \right\rangle \right\rangle $ but
rather for the ratio between this expectation value and the occupation
number $\left\langle \left\langle a_{k}^{\dagger }\left( t_{1}\right)
a_{k}\left( t_{1}\right) \right\rangle \right\rangle .$ To obtain an
expression for this ratio, we shall exploit the fact that we are seeking the
expectation value of an exponential. This expectation value is equivalent to
a generating functional for connected graphs $W$ where we couple $\xi
_{i}^{1}$ to a source $j_{1i}\left( t\right) =-\delta _{ik}\delta \left(
t-t_{1}\right) $ and $\xi _{l}^{2*}$ to $j_{2i}\left( t\right) =\delta
_{il}\delta _{p}\left( t-t_{1}\right) .$ Formally

\begin{equation}
\left\langle \left\langle a_{l}^{\dagger }\left( t_{1}\right) a_{k}\left(
t_{1}\right) \right\rangle \right\rangle =e^{iW\left[ j_{1i},j_{2i}\right] }
\label{id7}
\end{equation}
Decomposing $\xi _{i}^{a}$ in modes, we see that this is the same as
coupling the homogeneous terms to sources $j_{ai}^{0}=\mp \delta \left(
t-t_{1}\right) $ (independent of $k$ and $l$) and the inhomogeneous terms to
sources $\eta _{1p}^{\left( k\right) }\left( t\right) =-f_{pk}\delta \left(
t-t_{1}\right) $ and $\eta _{2p}^{\left( l\right) }\left( t\right)
=f_{pl}\delta \left( t-t_{1}\right) $.

In the diagonal case, we would have $k=l,$ and we would find the expectation
value on symmetry grounds alone. So now we aim to identify how the
nondiagonal case is different from the diagonal one. With this goal, we
write $\eta _{1p}^{\left( k\right) }\left( t\right) =\eta _{1p}^{\left(
l\right) }\left( t\right) +\delta \eta _{1p}^{\left( k,l\right) }\left(
t\right) ,$ $\delta \eta _{1p}^{\left( k,l\right) }\left( t\right) =\delta
\eta _{1p}^{\left( k,l\right) }\delta \left( t-t_{1}\right) $. $W$ has the
functional Taylor expansion

\begin{eqnarray}
W\left[ j_{a}^{0},\eta _{1p}^{\left( k\right) },\eta _{2p}^{\left( l\right)
}\right] &=&W\left[ j_{a}^{0},\eta _{1p}^{\left( l\right) },\eta
_{2p}^{\left( l\right) }\right]  \nonumber \\
&&-i\sum_{q>0}\frac{i^{q}}{q!}\sum_{p_{1}}...\sum_{p_{q}}G_{p_{1}...p_{q}}%
\left( t_{1},...t_{1}\right) \delta \eta _{1p_{1}}^{\left( k,l\right)
}...\delta \eta _{1p_{q}}^{\left( k,l\right) }  \label{id6}
\end{eqnarray}
where the $G_{p_{1}...p_{q}}\left( t_{1},...t_{1}\right) $ are the
time-ordered connected expectation values

\begin{equation}
G_{p_{1}...p_{q}}\left( t_{1},...t_{k}\right) =\left\langle \left\langle
X_{p_{1}}^{1}\left( t_{1}\right) ...X_{p_{q}}^{1}\left( t_{k}\right)
\right\rangle \right\rangle _{c}
\end{equation}
computed for a field driven by the sources $j_{a}^{0},\eta _{1p}^{\left(
l\right) },\eta _{2p}^{\left( l\right) }.$ Since the correlation functions
are continuous, and these sources turn on at the same time as the $\delta
\eta _{1p}^{\left( k\right) }$ they may be ignored. Keeping only up to
quadratic terms, we get

\begin{equation}
\left\langle \left\langle a_{l}^{\dagger }\left( t_{1}\right) a_{k}\left(
t_{1}\right) \right\rangle \right\rangle =e^{iW_{0}}\;\exp \left\{ \left(
\frac{-1}{2}\right) \sum_{p}\left\langle \left\langle
X_{p}X_{-p}\right\rangle \right\rangle \left| f_{pl}-f_{pk}\right|
^{2}\right\}
\end{equation}

\begin{equation}
W_{0}=W\left[ j_{a}^{0},\eta _{1p}^{\left( l\right) },\eta _{2p}^{\left(
l\right) }\right]
\end{equation}

Observe that we have separated the diagonal and non-diagonal contributions.
The former, encoded into $W_{0}$, shall be obtained below from the symmetry
condition Eq. (\ref{symmetry}), so we need not worry about explicitly
computing the path integral.

In the vacuum state

\begin{equation}
\left\langle \left\langle X_{p}X_{q}\right\rangle \right\rangle
=\left\langle \left\langle X_{p}^{*}X_{q}^{*}\right\rangle \right\rangle
=\delta _{p+q}\frac{U}{2\omega _{p}}  \label{propagator}
\end{equation}

\begin{equation}
\left| f_{pk}-f_{pl}\right| ^{2}=\frac{4}{N_{s}}\sin ^{2}\frac{\pi p\left(
k-l\right) }{N_{s}}
\end{equation}
so

\begin{equation}
\left\langle \left\langle a_{l}^{\dagger }\left( t_{1}\right) a_{k}\left(
t_{1}\right) \right\rangle \right\rangle =e^{iW_{0}}\;X\left[ l-k\right]
\end{equation}
where

\begin{equation}
X\left[ m\right] =\exp \left\{ \left( \frac{-2U}{N_{s}}\right) \sum_{p>0}%
\frac{1}{\omega _{p}}\sin ^{2}\frac{\pi pm}{N_{s}}\right\}  \label{sfdm}
\end{equation}

\subsection{Periodicity corrections}

We now consider the further corrections in Eqs. (\ref{key1}) and (\ref{key2}%
) coming from the periodicity conditions. For this, we need to evaluate $%
\left\langle \left\langle r_{j}^{\left( 2\right) }\left( 0\right)
A_{kl}\left( t_{1}\right) \right\rangle \right\rangle $ with

\begin{equation}
A_{kl}=\xi _{l}^{2*}-\xi _{k}^{1}
\end{equation}

First observe that there is no loss of generality in taking $t_{1}=0^{+}.$
Also there is no problem here with the homogeneous terms, because they
commute with $\alpha _{p}$ and so give a vanishing expectation value.

There is no loss of generality in setting the site index $k=0.$ In vacuum,

\begin{equation}
\left\langle \left\langle r_{j}^{\left( 2\right) }\left( 0\right)
A_{0m}\right\rangle \right\rangle =\frac{i}{2}\left[ \delta _{0j}-\delta
_{mj}-\frac{1}{n}\left[ M_{jm}+M_{j0}\right] \right] \equiv iv_{j}^{m}
\label{vector}
\end{equation}
By symmetry, we must have $\sigma _{100}=n.$ We use this condition to
determine $W_{0}$, getting

\begin{equation}
\sigma _{1m0}=nX\left[ m\right] \frac{Y\left[ m\right] }{Y\left[ 0\right] }
\label{dm}
\end{equation}
where $X\left[ m\right] $ is defined in Eq. (\ref{sfdm}) and

\begin{equation}
Y\left[ m\right] =\sum_{k_{i}}e^{-2\pi \left[ i\sum_{i}k_{i}v_{i}^{m}+\pi
\sum_{ij}k_{i}M_{ij}k_{j}\right] }  \label{yofm}
\end{equation}
where $v_{j}^{m}$ was introduced in Eq. (\ref{vector}).

These expressions determine the one-body density matrix in the two limiting
cases $U/J\rightarrow 0$ and $U/J\rightarrow \infty .$ In the former,
corresponding to the deep superfluid phase, we have $Y\left[
m\right]/Y\left[ 0\right] \sim 1$ and $\left. \sigma _{10m}\right| _{U/J\ll
1}=nX\left[ m\right] $, given by Eq. (\ref{sfdm}).

In the latter limit, corresponding to the deep Mott insulator phase, we may
replace the sum in Eq. (\ref{yofm}) by an integral over continuous variables
$k_{i}$. In this continuum approximation we get

\begin{equation}
\left. \sigma _{1m0}\right| _{U/J\gg 1}=n\;\left( X\left[ m\right] \right)
^{2}  \label{midm}
\end{equation}

In the intermediate region, we may approximate Eq. (\ref{yofm}) by

\begin{equation}
Y\left[ m\right] =\prod_{p>0}\vartheta _{3}\left( \pi C_{p}^{m},e^{-2\pi
^{2}E_{p}}\right) \vartheta _{3}\left( \pi D_{p}^{m},e^{-2\pi
^{2}F_{p}}\right)  \label{approxy}
\end{equation}
where $\vartheta _{3}$ is the Elliptic Theta function \cite{WW02}

\begin{equation}
C_{p}^{m}=\frac{1}{N_{s}}\left[ 1+2\sum_{l>0}\left| \cos \left[ \frac{2\pi pl%
}{N_{s}}\right] \right| \right] \left\{ \sin ^{2}\left[ \frac{\pi pm}{N_{s}}%
\right] -\frac{\nu _{p}}{2\omega _{p}}\cos ^{2}\left[ \frac{\pi pm}{N_{s}}%
\right] \right\}  \label{c}
\end{equation}

\begin{equation}
D_{p}^{m}=\frac{1}{N_{s}}\left\{ \sum_{l>0}\left| \sin \left[ \frac{2\pi pl}{%
N_{s}}\right] \right| \right\} \left[ 1+\frac{\nu _{p}}{2\omega _{p}}\right]
\sin \left[ \frac{2\pi pm}{N_{s}}\right]  \label{d}
\end{equation}

\begin{equation}
E_{p}=\frac{n\nu _{p}}{4N\omega _{p}}\left\{ 1+2\sum_{l>0}\left| \cos \left[
\frac{2\pi pl}{N_{s}}\right] \right| \right\} ^{2}  \label{e}
\end{equation}

\begin{equation}
F_{p}=\frac{n\nu _{p}}{N_{s}\omega _{p}}\left\{ \sum_{l>0}\left| \sin \left[
\frac{2\pi pl}{N_{s}}\right] \right| \right\} ^{2}  \label{f}
\end{equation}

This approximation is further discussed in Appendix B. We have checked its
accuracy for small lattices, by comparing it to a numerical evaluation of
Eq. (\ref{yofm}).

\section{Results and Remarks}

\subsection{Results}

In this Section, we shall compare the analytic results above against an
exact calculation of the momentum distribution function, Eq. (\ref{mdf}),
for an one dimensional lattice of $9$ sites and $9$ atoms ($n=1$). The exact
solution was obtained by numerical diagonalization of the Bose Hubbard
Hamiltonian. We set $J=1$ and change $U$ from $0$ to $60$. We have performed
similar calculations for $5$ and $7$ sites, finding the results to be
totally consistent with the $N=9$ case. The allowed values of momentum are
given by $p\left( q\right) =q\left( 2\pi \hbar /aN_{s}\right) $ $,$ where $a$
is the lattice spacing and $q$ is an integer. Since by symmetry $p\left(
-q\right) =p\left( q\right) $, there are only five independent occupation
numbers, corresponding to $q=0$ (the condensate) to $4.$ These are plotted
in Figs. 1 to 5, respectively.

In these figures we have also plotted the occupation numbers as given by the
PNC method ( Bogoliubov) calculations , and by first order strong coupling
perturbation theory. In all the plots,The solid-red line is our prediction,
the black dash-dotted line is the exact numerical solution, the pink dots
correspond to first order strong coupling perturbation theory, and the
blue-dashed line to the PNC method.

We see that for these small lattices our model fares worse than perturbation
theory or the PNC approach in the corresponding limits of the deep Mott or
superfluid regions, but unlike these formalisms, it sustains an uniform
accuracy throughout. It therefore achieves the goals set in the Introduction.

\begin{figure}[tbh]
\leavevmode {\includegraphics[width=4. in]{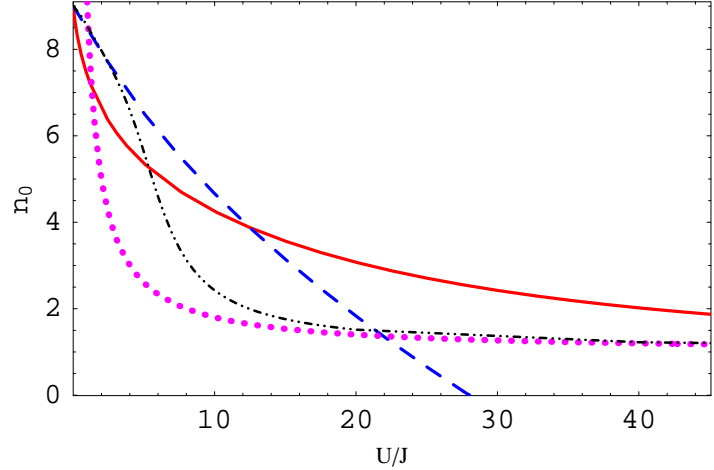}} \
\caption{The occupation number for the homogeneous mode, as a function of $U$%
; $n=1$, $N_s=9$ and $J=1$. The solid-red line is our prediction, the black
dash-dotted line is the exact numerical solution. The pink dots correspond
to first order strong coupling perturbation theory, and the blue-dashed line
to the PNC method.}
\label{Fig1}
\end{figure}

\begin{figure}[tbh]
\leavevmode {\includegraphics[width=4. in]{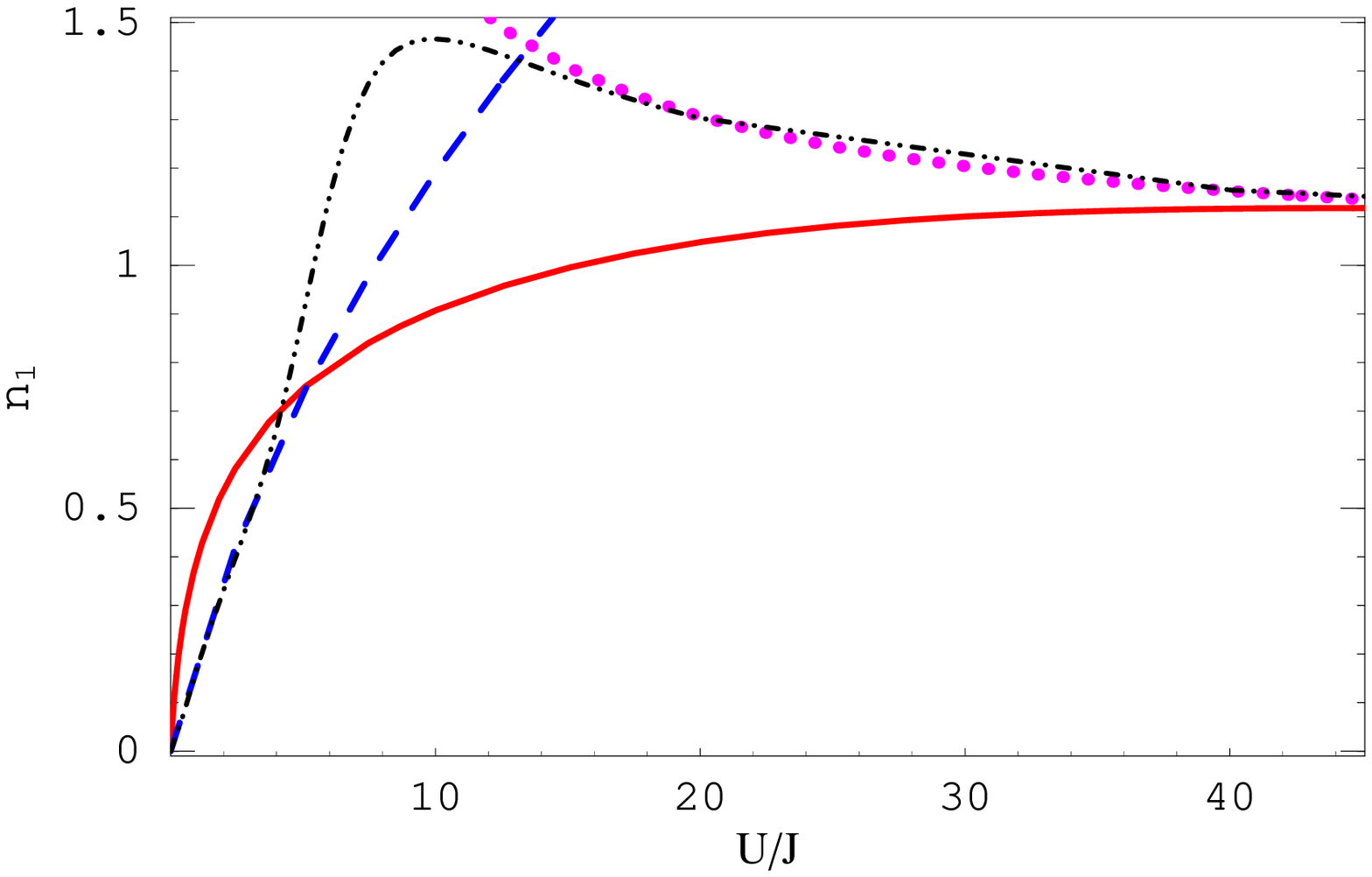}} \
\caption{The occupation number for the first mode as a function of $U$; the
conventions are the same as in Fig. 1.}
\label{Fig2}
\end{figure}

\begin{figure}[tbh]
\leavevmode {\includegraphics[width=4. in]{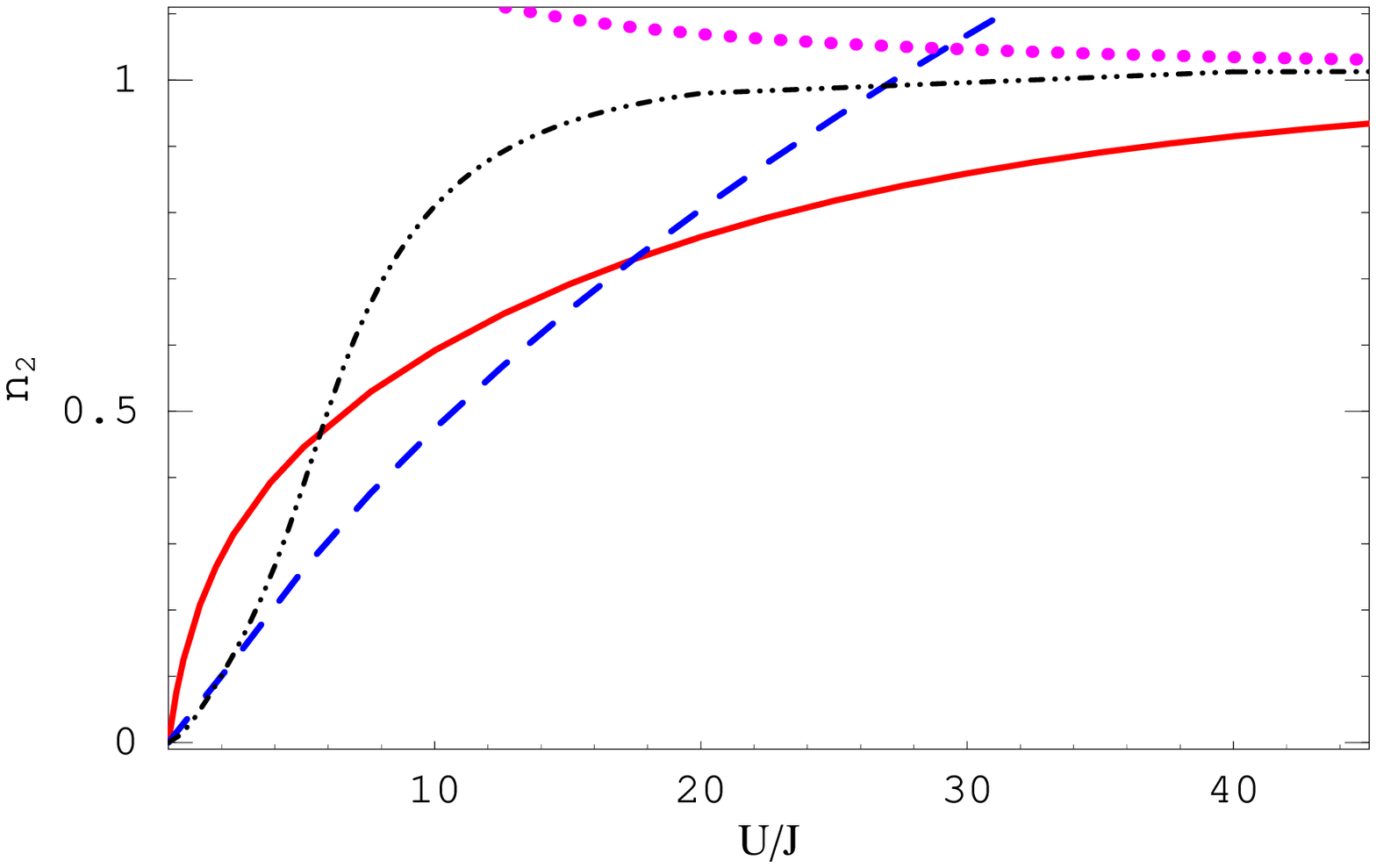}} \
\caption{The occupation number for the second mode as a function of $U$; the
conventions are the same as in Fig. 1.}
\label{Fig3}
\end{figure}

\begin{figure}[tbh]
\leavevmode {\includegraphics[width=4. in]{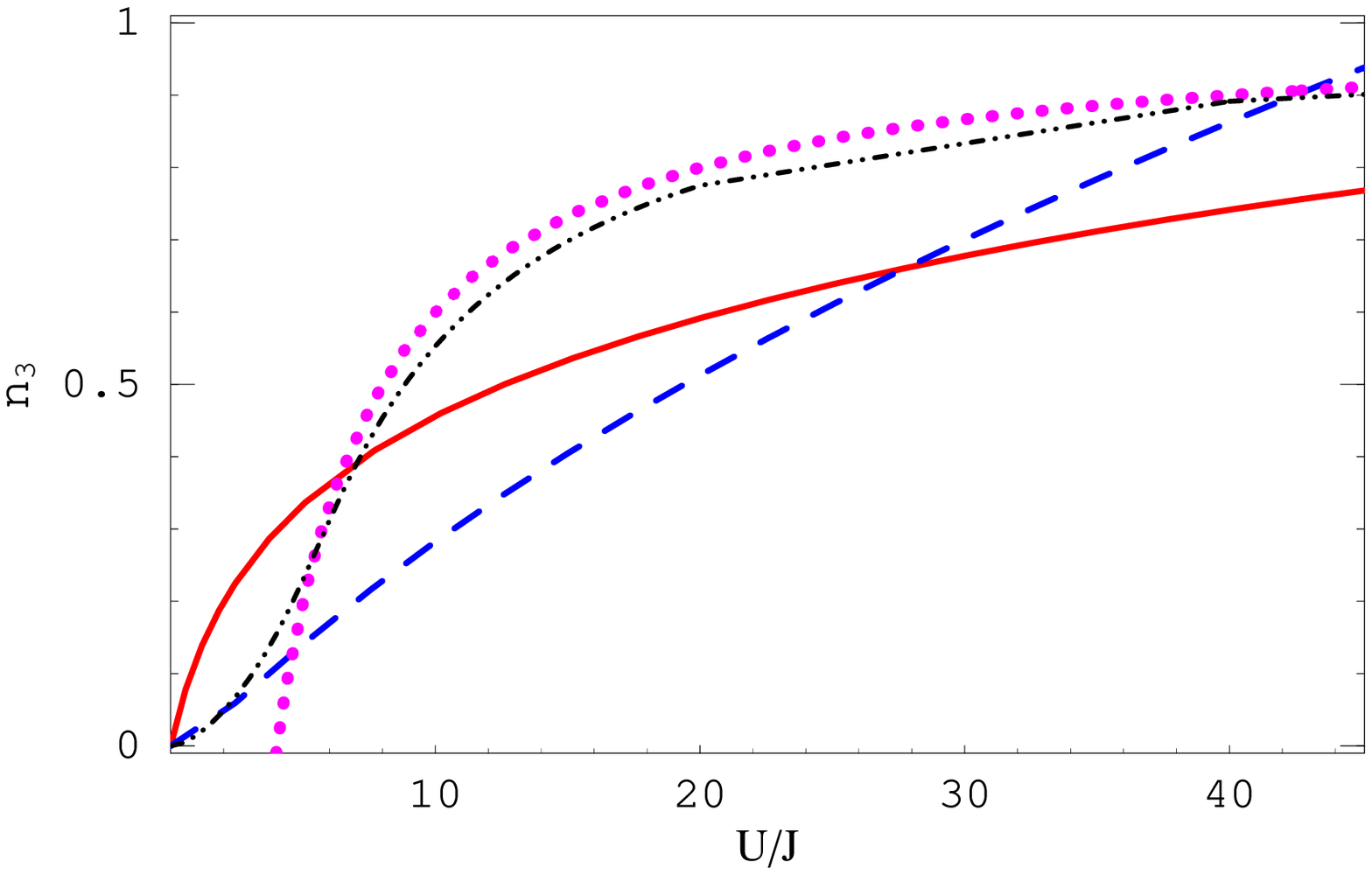}} \
\caption{The occupation number for the third mode as a function of $U$; the
conventions are the same as in Fig. 1.}
\label{Fig4}
\end{figure}

\begin{figure}[tbh]
\leavevmode {\includegraphics[width=4. in]{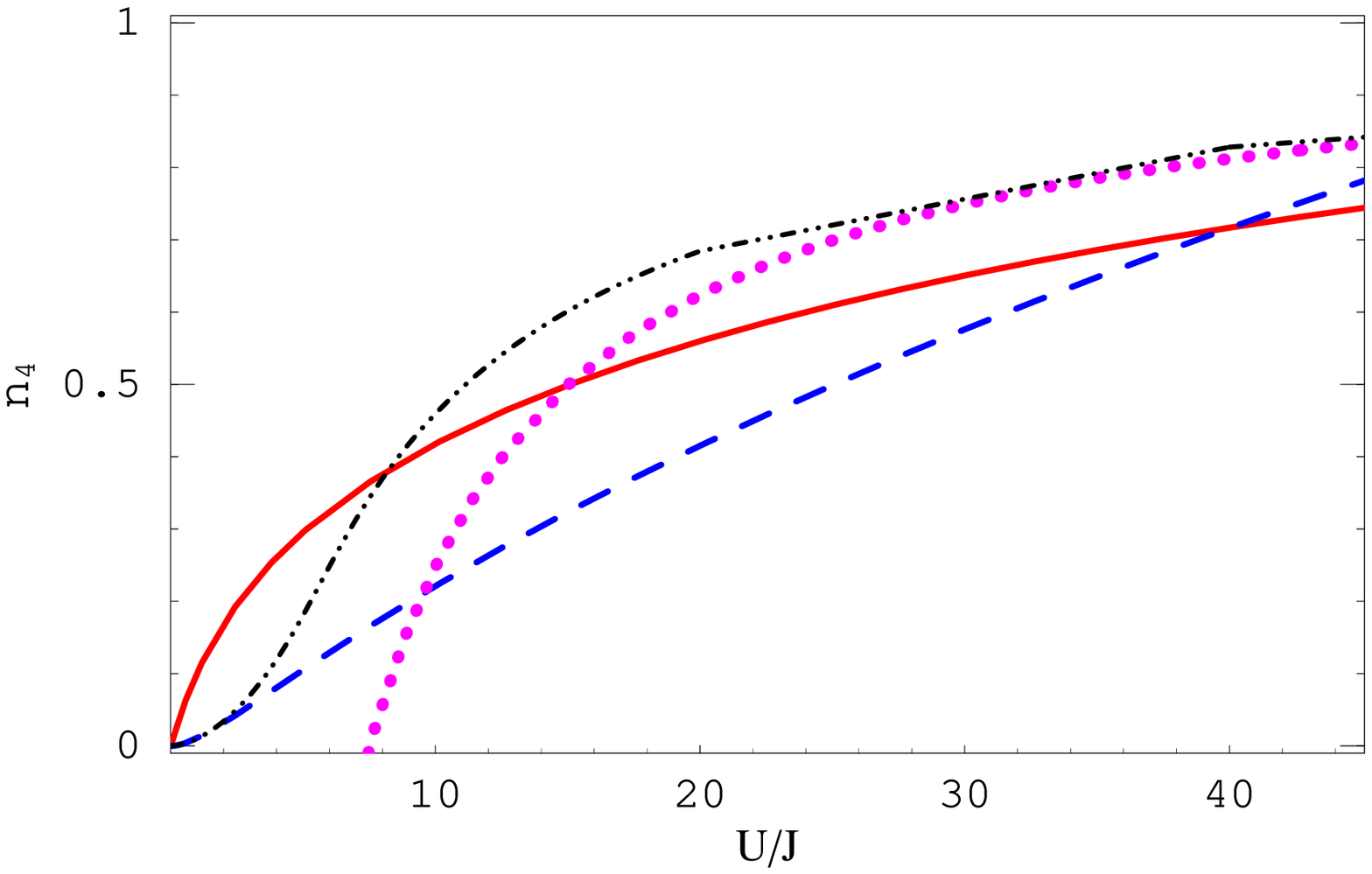}} \
\caption{The occupation number for the fourth mode as a function of $U$; the
conventions are the same as in Fig. 1.}
\label{Fig5}
\end{figure}

\begin{figure}[tbh]
\leavevmode {\includegraphics[width=4. in]{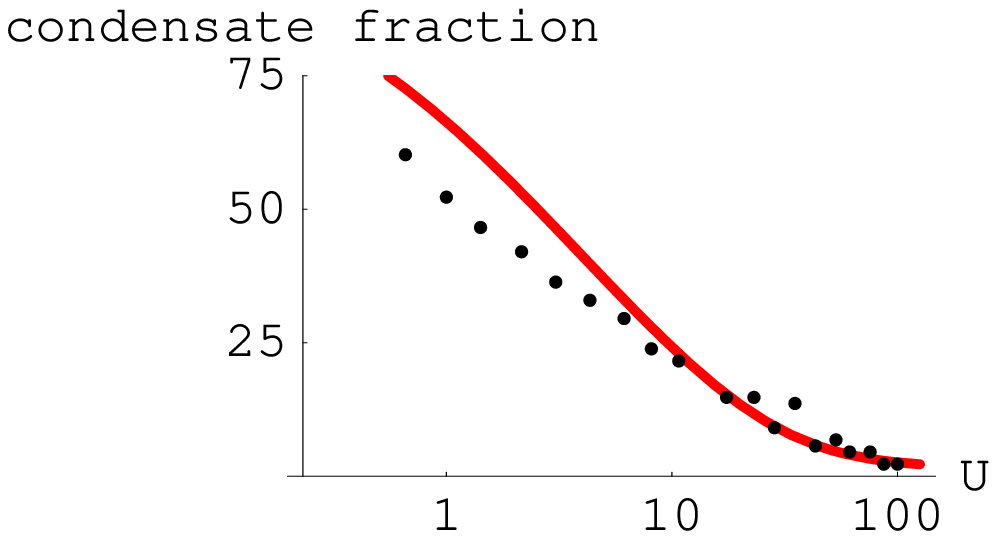}} \
\caption{ Condensate fraction ($\%$) plotted against $U$. Red solid line:
prediction from our model using the parameters $n=1$, $N_{s}=61$ and $J=1$,
Dots: Experimental points obtained from Fig. 4a in Ref \protect\cite{SMSKE04}%
.}
\label{Fig6}
\end{figure}

\subsection{Remarks}

In this paper we have presented an analytic approximation for the one body
density matrix (or equivalently, its Fourier transform, the momentum
distribution) for a cold gas of structureless bosons in an homogeneous
optical lattice. We have focused on the regime of low integer filling factor
near the superfluid - insulator transition, which is not sufficiently
covered in the literature. We have checked our results against exact
calculations for small lattices, and against the theoretical predictions
from the Bogoliubov approach and first order strong coupling perturbation
theory. Our model interpolates between these theoretical alternatives,
keeping an uniform accuracy in the transition region.

Our model works deep in the MI region, because it is built to be exact when $%
J=0$. This is an advantage of our choice of variables over the usual density
-phase variables. In the superfluid regime the model predicts that quantum
fluctuations in the $X_{p}$ degrees of freedom scale like $U$ (cf. Eq. (\ref
{propagator}) and thus also qualitatively captures the decay of condensate
population and the increase of non-condensate atoms.

However, the agreement is not perfect. The qualitative but not quantitative
agreement suggest that higher order corrections are required for a proper
description of the physics. For observables like the number fluctuations at
one site, which vanish in the Mott regime according to the linearized
approximation, for example, higher order corrections would be dominant.

Quantum corrections will also be important for the dynamic structure factor.
There is no contradiction between the phonon spectrum of our model (cfr. Eq.
(\ref{phonspec})) and a gapped dynamic structure factor, because in the Mott
regime the amplitudes of the single phonon poles go to zero, while other
poles arise because of higher order corrections. However, in this paper, we
have not presented actual results for the particle number fluctuations nor
the dynamic structure factor; these must be included in the list of
unfinished business we discussed in the introduction.

A preliminary comparison we made against available experimental
results \cite {SMSKE04,SSMKE04} of the condensate fraction from an
array of one-dimensional lattices contained within a three
dimensional trap for variable $U/J$ showed fair agreement between
the experimental results and the predictions of our model. In these
experiments, the central tubes had around $N_{s}=60$ populated sites
\cite{SSMKE04}. The mean occupation number was close to $n=2$ near
the center of the trap, and close to $n=1$ if averaged over all
lattices \cite{K05}. We have compared the experimental results to
the predictions of our model for several values of $N_{s}$ around
$60,$ and filling fractions $n=1$ and $2.$ The results are fairly
independent of $N_{s}$ in this range, and very sensitive to $n$
instead.  As a typical representative, we show in Fig. 6 the
prediction of our model for the condensate fraction for $N_{s}=61$
and $n=1.$We have superimposed the experimental results as reported
in \cite{SMSKE04}.

We do not regard this as a validation of our model, since it was derived for
a translationally invariant lattice and the parabolic confinement is not
adequately included in our model. Nevertheless, the agreement is encouraging
and suggest that our model might be more suitable for trapped systems as in
this case, in contrast to the commensurate translationally invariant
lattice, there is not a sharp MI transition. We defer a detailed discussion
to a future communication \cite{RHC05}.

In summary, in our view, the most important contribution of this work is
that it is the first step in the formulation of a quantum field theoretical
approach capable of dealing with the intermediate regime. Even though the
agreement with exact numerical solutions is not perfect, we find it
satisfactory because we are using only the first order approximation. It is
a reasonable expectation that by including higher order corrections we might
narrow the present gap. We are perhaps still a long way from a reliable,
fully nonequilibrium model of the initialization process of a QIP device
based on cold atoms on an optical lattice, but from this work we have gained
some confidence that we are moving in the right direction.\newline

\noindent\textbf{Acknowledgments} EC acknowledges support from Universidad
de Buenos Aires, CONICET and ANPCyT (Argentina); BLH from NSF grant
PHY-0426696.

A.M. Rey is supported from an the Advanced Research and Development
Activity (ARDA) contract and the U.S. National Science Foundation
through a  grant PHY-0100767 and a  grant from  the Institute of
Theoretical, Atomic, Molecular  and Optical Physics at Harvard
University and Smithsonian Astrophysical observatory.

We thank M. K\"{o}hl for pointing out the relevance of refs. \cite
{SMSKE04,SSMKE04} to this work and for discussing details of the experiment
with us, and V. Penna, M. Martin-Delgado and M. Rigol for useful comments.

\appendix

\section{Approximate approaches to the momentum distribution function}

In this appendix we shall derive the formulae we plotted in the Figures to
match against our model. We include it only to dispell any ambiguity
regarding notation.

\subsection{Strong coupling Rayleigh - Schrodinger perturbation theory}

This is just ordinary perturbation theory in the parameter $J,$ starting
from the state $\left| MI\right\rangle $ in Eq. (\ref{mottstate}). The BH
Hamiltonian Eq. (\ref{nbodyh}) is written as $H=H_{0}+H_{1},$ where $H_{0}$
is the $U$ term and $H_{1}=-\sum_{ij}J_{ij}a_{i}^{\dagger }a_{j}.$ Since $%
\left\langle MI\left| H_{1}\right| MI\right\rangle =0$, the vacuum energy is
unchanged to first order. $H_{1}\left| MI\right\rangle $ is a superposition
of one particle-hole states, all of which have energy $U$ above the vacuum,
so the first order ground state is

\begin{equation}
\left| T\right\rangle =\left| MI\right\rangle -\frac{1}{U}H_{1}\left|
MI\right\rangle
\end{equation}
and the momentum distribution function is

\begin{equation}
n_{q}=n+\frac{4J}{U}n\left( n+1\right) \cos \left[ 2\pi \frac{q}{N_{s}}%
\right]
\end{equation}

\subsection{The PNC method}

To simplify the problem, we shall consider only the case of a homogeneous,
time - independent lattice.

The starting point of the method is the Heisenberg equation of motion for
the destruction operator $a_{j}$

\begin{equation}
\left( -i\right) \frac{\partial }{\partial t}a_{j}\left( t\right) =\left[
H,a_{j}\left( t\right) \right] =\sum_{i}J_{ij}a_{i}-Ua_{j}^{\dagger
}a_{j}^{2},  \label{heisenberg}
\end{equation}

Parameterize

\begin{equation}
a_{j}=\frac{e^{-i\mu t}}{\sqrt{N_{s}}}a_{0}\left[ 1+\sum_{p\neq 0}e^{2\pi
ipj/N_{s}}\Lambda _{p}\right]
\end{equation}

There are three key observations: 1) the operators $\Lambda _{p}$ preserve
total particle number, 2) the fact that the one-body density matrix allows
for a homogeneous eigenvector implies $\left\langle a_{0}^{\dagger
}a_{0}\Lambda _{p}\right\rangle =0$ for all $p$, and 3) we have the exact
identity

\begin{equation}
a_{0}^{\dagger }a_{0}\left[ 1+\sum_{p\neq 0}\Lambda _{p}^{\dagger }\Lambda
_{p}\right] =N
\end{equation}

Now we develop a perturbative expansion in inverse powers of $N,$ assuming $%
a_{0}\sim O\left( \sqrt{N}\right) $ and $\Lambda _{p}\sim O\left( 1/\sqrt{N}%
\right) .$ Multiplying Eq. (\ref{heisenberg}) by $a_{0}^{\dagger }$ we get,
to first order

\begin{equation}
\mu =Un-2J
\end{equation}

\begin{equation}
i\frac{d\Lambda _p}{dt}=\frac{\nu _p}2\Lambda _p+Un\left( \Lambda _p+\Lambda
_{-p}^{\dagger }\right)
\end{equation}

As usual, we seek a solution through a Bogoliubov transformation. Taking
into account the commutation relations for the $\Lambda _{p}$ we get
\begin{equation}
\Lambda _{p}=\frac{1}{\sqrt{N}}\left\{ e^{-i\omega
_{p}t}c_{p}A_{p}+e^{i\omega _{p}t}s_{p}A_{-p}^{\dagger }\right\}
\end{equation}

\begin{equation}
c_{p}^{2}-s_{p}^{2}=1
\end{equation}
(in this simple problem, we may assume the Bogoliubov coefficients are
real). We get

\begin{equation}
\left[ \omega _{p}-Un-\frac{\nu _{p}}{2}\right] c_{p}-Uns_{p}=0
\end{equation}

\begin{equation}
Unc_{p}+\left[ \omega _{p}+Un+\frac{\nu _{p}}{2}\right] s_{p}=0
\end{equation}
from where we recover the dispersion relation Eq. (\ref{phonspec}) and

\begin{equation}
s_{p}=-\frac{1}{\sqrt{2}}\left[ \frac{Un+\frac{\nu _{p}}{2}}{\omega _{p}}%
-1\right] ^{1/2}
\end{equation}

The momentum distribution function is $n_p=s_p^2$ for $p\neq 0$, and $%
n_0=N-\sum_{p\neq 0}n_p$ for the homogeneous mode.

\section{Approximate formula for the one-body density matrix}

The idea is to evaluate Eq. (\ref{yofm}) by decomposing the quadratic term
in a sum of squares. Of course, one possibility is to write

\begin{equation}
k_{j}=\frac{1}{\sqrt{N_{s}}}\sum_{p}e^{2\pi ipj/N_{s}}\widetilde{k}_{p}
\end{equation}

The problem is that the requirement that all $k_{j}$ be integer places a
highly nontrivial constraint on the $\widetilde{k}_{p}.$

Consider instead the functions

\begin{equation}
f_{p}\left( j\right) =\mathrm{sign}\left[ \cos \left[ \frac{2\pi pj}{N_{s}}%
\right] \right]
\end{equation}

\begin{equation}
g_{p}\left( j\right) =\mathrm{sign}\left[ \sin \left[ \frac{2\pi pj}{N_{s}}%
\right] \right]
\end{equation}

These functions are not orthogonal, but they are a basis. Therefore we can
always write

\begin{equation}
k_{j}=a_{0}+\frac{1}{2}\sum_{p>0}\left( a_{p}f_{p}\left( j\right)
+b_{p}g_{p}\left( j\right) \right)  \label{app1}
\end{equation}

Observe that $f_{0}$ is always a null eigenvector of $M_{ij}.$ We expect the
$f$ and $g^{\prime }$s will be approximate eigenvectors. For a large number
of sites, $e^{2\pi iqj/N_{s}}$ and $f_{p},g_{p}$ will be nearly orthogonal
unless $q=\pm p,$ and we shall have

\begin{equation}
f_{p}\left( j\right) \sim \frac{2}{N_{s}}\left\{ 1+2\sum_{l>0}\left| \cos
\left[ \frac{2\pi pl}{N_{s}}\right] \right| \right\} \cos \left[ \frac{2\pi
pj}{N_{s}}\right]
\end{equation}

\begin{equation}
g_{p}\left( j\right) \sim \frac{4}{N_{s}}\left\{ \sum_{l>0}\left| \sin
\left[ \frac{2\pi pl}{N_{s}}\right] \right| \right\} \sin \left[ \frac{2\pi
pj}{N_{s}}\right]
\end{equation}
So

\begin{equation}
\sum_{l}M_{jl}f_{p}\left( l\right) \sim A_{p}f_{p}\left( j\right)
\end{equation}

\begin{equation}
\sum_{l}M_{jl}g_{p}\left( l\right) \sim A_{p}g_{p}\left( j\right)
\end{equation}
where of course $A_{0}=0$ and

\begin{equation}
A_{p}=\frac{n\nu _{p}}{2\omega _{p}}\qquad p\neq 0
\end{equation}

Then from the decomposition Eq. (\ref{app1}) we get

\begin{equation}
\sum_{i}k_{i}v_{i}^{m}=\sum_{p>0}\left( a_{p}C_{p}^{m}-b_{p}D_{p}^{m}\right)
\end{equation}

\begin{equation}
\sum_{ij}k_{i}M_{ij}k_{j}=\sum_{p>0}\left(
a_{p}^{2}E_{p}+b_{p}^{2}F_{p}\right)
\end{equation}
where the coefficients are given in Eqs. (\ref{c}), (\ref{d}), (\ref{e}) and
(\ref{f}).

Although each term of the series Eq. (\ref{yofm}) factorizes, there are
correlations among the $a$ \ and $b$ coefficients from the discreteness of
the $k_{j}.$ For example, for three sites we have that $b_{1}=k_{1}-k_{-1}$  must be an integer, but $a_{1}=k_{0}-k_{1}+(b_{1}/2)$ will be integer if $b_{1}$ is even or
half-integer if $b_{1}$ is odd. However, when the number of sites is large
we may neglect these correlations, and assume that the $a$ and $b$
coefficients simply take integer values. Under this approximation, the
multiple sum Eq. (\ref{yofm}) factorizes, and we obtain Eq. (\ref{approxy}),
where $\vartheta _{3}$ is the Elliptic Theta function \cite{WW02}

\begin{equation}
\vartheta _{3}\left( z,q\right) =1+2\sum_{n=1}^{\infty }q^{n^{2}}\cos \left[
2nz\right]
\end{equation}


\end{document}